\documentclass{pasa}%

\usepackage{graphicx}
\usepackage{amsmath}	
\usepackage{amssymb}	

\usepackage{fixfoot}
\usepackage[clockwise, figuresright]{rotating}
\usepackage{color,soul}
\usepackage{multirow}
\usepackage{IEEEtrantools}
\usepackage{bm}
\usepackage{textcomp}
\usepackage{xcolor}
\usepackage{calligra}
\usepackage{dashrule}
\usepackage{scalerel}
\usepackage{footmisc}
\usepackage{xspace}

\newcommand{\farcs}{\mbox{\ensuremath{.\!\!^{\prime\prime}}}}
\newcommand{\fdg}{\mbox{\ensuremath{.\!\!^\circ}}}
\newcommand{\degr}{\ensuremath{^{\circ}}}
\newcommand{\micron}{\ensuremath{\mu}{\rm m}}
\newcommand{\arcmin}{\ensuremath{^\prime}}
\newcommand{\arcsec}{\ensuremath{^{\prime\prime}}}
\newcommand{\fs}{\ensuremath{.\!\!^{\rm s}}}

\definecolor{MATLAB_blue}{rgb}{0,0.4470,0.7410}
\definecolor{MATLAB_red}{rgb}{0.8500,0.3250,0.0980}
\definecolor{MATLAB_orange}{rgb}{0.9290,0.6940,0.1250}
\definecolor{MATLAB_purple}{rgb}{0.4940,0.1840,0.5560}
\definecolor{MATLAB_green}{rgb}{0.4660,0.6740,0.1880}
\definecolor{MATLAB_cyan}{rgb}{0.3010,0.7450,0.9330}
\definecolor{MATLAB_maroon}{rgb}{0.6350,0.0780,0.1840}

\definecolor{LimeGreen}{rgb}{0.1961,0.8039,0.1961}
\definecolor{Orange}{rgb}{1,0.6471,0}

\title[Publications of the Astronomical Society of Australia]{Refining the mass estimate for the intermediate-mass black hole candidate in NGC~3319}

\author[Davis \& Graham]{
Benjamin L. Davis$^{1,2,}$\thanks{Author for correspondence: BLD, E-mail: \href{mailto:ben.davis@nyu.edu}{ben.davis@nyu.edu}}\,\,\href{https://orcid.org/0000-0002-4306-5950}{\hskip2pt\includegraphics[width=9pt]{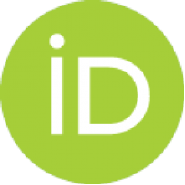}}
and Alister W. Graham$^{1}$\href{https://orcid.org/0000-0002-6496-9414}{\hskip2pt\includegraphics[width=9pt]{orcid-ID.pdf}}
\affil{$^1$Centre for Astrophysics and Supercomputing, Swinburne University of Technology, Hawthorn, VIC 3122, Australia}%
\affil{$^2$Center for Astro, Particle, and Planetary Physics (CAP$^3$), New York University Abu Dhabi}
}%

\jid{PASA}
\doi{10.1017/pas.\the\year.xxx}
\jyear{\the\year}

\usepackage{aas_macros}
\usepackage{hyperref} 
\hypersetup{colorlinks,citecolor=blue,linkcolor=blue,urlcolor=blue}

\begin{document}
\begin{frontmatter}
\maketitle

\begin{abstract}
Recent X-ray observations by Jiang et al.\ have identified an active galactic nucleus (AGN) in the bulgeless spiral galaxy NGC~3319, located just $14.3\pm1.1$\,Mpc away, and suggest the presence of an intermediate-mass black hole (IMBH; $10^2\leq M_\bullet/\mathrm{M_{\odot}}\leq10^5$) if the Eddington ratios are as high as 3 to $3\times10^{-3}$. In an effort to refine the black hole mass for this (currently) rare class of object, we have explored multiple black hole mass scaling relations, such as those involving the (not previously used) velocity dispersion, logarithmic spiral-arm pitch angle, total galaxy stellar mass, nuclear star cluster mass, rotational velocity, and colour of NGC~3319, to obtain ten mass estimates, of differing accuracy. We have calculated a mass of $3.14_{-2.20}^{+7.02}\times10^4\,\mathrm{M_\odot}$, with a confidence of 84\% that it is $\leq$$10^5\,\mathrm{M_\odot}$, based on the combined probability density function from seven of these individual estimates. Our conservative approach excluded two black hole mass estimates (via the nuclear star cluster mass, and the fundamental plane of black hole activity --- which only applies to black holes with low accretion rates) that were upper limits of $\sim$$10^5\,{\rm M}_{\odot}$, and it did not use the $M_\bullet$--$L_{\rm 2-10\,keV}$ relation's prediction of $\sim$$10^5\,{\rm M}_{\odot}$. This target provides an exceptional opportunity to study an IMBH in AGN mode and advance our demographic knowledge of black holes. Furthermore, we introduce our novel method of meta-analysis as a beneficial technique for identifying new IMBH candidates by quantifying the probability that a galaxy possesses an IMBH.
\end{abstract}

\begin{keywords}
black hole physics -- galaxies: active -- galaxies: evolution -- galaxies: individual: NGC~3319 -- galaxies: spiral -- galaxies: structure\\
\\
Original unedited manuscript, accepted for publication by PASA, May 7, 2021.
\end{keywords}
\end{frontmatter}

\section{Introduction}

There is a largely-missing population of intermediate-mass black holes (IMBHs)
with masses higher than those formed by stable, single stars today ($M_\bullet\lesssim100\,\mathrm{M_{\odot}}$) and less massive than the supermassive black holes
(SMBHs; $10^5\,\mathrm{M_{\odot}}\leq M_\bullet\lesssim10^{10}\,\mathrm{M_{\odot}}$)\footnote{The massive central object in the quasar TON~618 is alleged to have the most massive black hole with a mass of $6.61\times10^{10}\,\mathrm{M_{\odot}}$, estimated from its H$\beta$ emission line and a virial $f$-factor of 5.5 \citep{Shemmer:2004,Onken:2004}.} known to reside at the centres of
massive galaxies.  Not surprisingly, astronomers around the world have been hotly
pursuing the much-anticipated discovery of IMBHs for some time \citep[e.g.][]{Miller:2004}. 
In addition to 
providing a fundamental input to 
the cosmic inventory of our Universe, the abundance, or rarity, of IMBHs
has implications for the formation of the Universe's SMBHs \citep{Graham:2016c,Mezcua:2017,Koliopanos:2017b,Inayoshi:2019,Sahu:2019}.

As yet, there is no consensus as to how SMBHs came to be.  While the
observed extent of quasar activity over the history of our Universe has
revealed that the accretion of baryons fattened them up \citep[e.g.][]{Soltan:1982,Shankar:2004}, we do not know what their (potentially range of)
birth masses were.  Some theories have speculated that their birth or `seed'
masses were $\approx$$10^5\,{\rm M}_{\odot}$, thereby providing a kick-start to
explain the early-formation of the high-$z$, active
galactic nuclei (AGN) with sizeable black hole masses around $\approx$$10^9\,{\rm
  M}_{\odot}$ \citep[e.g.][]{Mortlock:2011,Yang:2020,Mignoli:2020}. Theories have included primordial black holes \citep[e.g.][]{Grobov:2011}, massive metal-free Population III stars which
subsequently collapse \citep[or collide, e.g.][]{Seguel:2019} to form massive black holes \citep[e.g.][]{Madau:2001,Schneider:2002}, or the direct collapse of
massive gas clouds, effectively by-passing the stellar phase of evolution
\citep[e.g.][]{Bromm:2003,Mayer:2010}.

The suggestion of massive seeds arose from the notion that the `Eddington
limit' \citep{Eddington:1925} of gas accretion onto a black hole implied that stellar-mass black holes did not have
sufficient time to grow into the SMBHs observed in the young, 
high-redshift AGN.  
However, the Eddington limit on the accretion
rate applies only to (unrealistic) spherical conditions \citep{Nayakshin:2012,Alexander:2014} and
can be significantly exceeded in real systems. For example, super-critical (super-Eddington) accretion flows onto massive black holes can occur when the accretion flow is mostly confined to the disk
  plane while most of the radiation emerges in outflows along the rotation axis \citep{Abramowicz:1980,Jiang:2014,Pezzulli:2016}. Hyper-Eddington accretion rates can exist in spherically-symmetric accretion flows when energy advection reduces radiative efficiency \citep{Inayoshi:2016}. Thus, the practicality of super-critical accretion has been invoked to explain the early existence of SMBHs at high redshifts \citep{Volonteri:2005,Volonteri:2012,Volonteri:2012b,Volonteri:2015}. Besides, most ultra-luminous X-ray sources
are nowadays explained as stellar-mass X-ray binaries accreting much faster than their Eddington limit \citep{Feng:2011,Kaaret:2017}. Such accretion negates
the need for massive black hole seeds.

An additional motive for starting AGN with massive seeds was that black holes with
masses intermediate between that of stellar-mass black holes
and SMBHs had not been directly 
observed, and therefore seemed not to exist.  However, this may be a
sample selection bias because the sphere-of-gravitational-influence around
such IMBHs, where one would directly observe a Keplerian 
rotation curve, is typically too small to resolve spatially.
Furthermore, there is now a rapidly rising number of
IMBH candidates based upon indirect estimates of the black hole mass \citep{Farrell:2009,Secrest:2012,Baldassare:2015,Graham:2016b,Baumgardt:2017,Nguyen:2017,Chilingarian:2018,Mezcua:2018,Jiang:2018,Nguyen:2019,Graham:2019a,Graham:2019b,Woo:2019,Lin:2020}. In addition, there are currently five IMBH candidates in the Milky Way \citep{Takekawa:2020}.

There is no shortage of scenarios for how a bridging population of IMBHs may have 
arisen. Possible pathways include the runaway collapse of dense `nuclear star clusters' \citep{Portegies_Zwart:2002,Davies:2011,Lupi:2014,Stone:2017}, especially if gas-drag and dynamical friction are in play at the centre of a galaxy, or the gas-fuelled growth of a stellar-mass black hole that has not yet
devoured enough material to become an SMBH \citep{Natarajan:2020}. These ideas would place, at least some, IMBHs at the centres of galaxies, where established black hole mass scaling relations involving some property of the host galaxy can be applied.

Recent \textit{Chandra X-ray Observatory} \citep[\textit{CXO};][]{Weisskopf:2000} observations \citep[][see also \citealt{Chilingarian:2018} and \citealt{Bi:2020}]{Soria:2016}, have discovered IMBH candidates at the centres\footnote{Some of the off-centre X-ray sources
that were detected may also be IMBHs. Indeed, the best localised IMBH candidate to date
is an off-centre source in the galaxy ESO 243-49 \citep{Farrell:2009},
whose optical counterpart was discovered by \citet{Soria:2010} and is thought to be the nucleus of an in-falling galaxy.  However,
the likelihoods of these off-centre targets
being IMBHs are generally considered to be notably lower than that of the
central targets --- although perhaps not zero \citep[e.g.][]{Barrows:2019,Bellovary:2021}.} of
several nearby, low-mass galaxies.
Long exposures have enabled the discovery of faint
X-ray point-sources (consistent with low-mass black holes accreting with low Eddington ratios) in galaxies which have been predicted to host a central IMBH
based upon each galaxy's velocity dispersion, luminosity, and spiral-arm pitch angle \citep{Koliopanos:2017,Graham:2019a,Graham:2019b}. The high-energy X-ray photons, originating from the (not so) dead centres of the galaxies, are likely coming from the accretion disks around black holes because of their point-source nature, where emission favours active black holes rather than spatially extended star formation.

Several studies have identified IMBH candidates in galaxies based on single, or a few, black hole mass estimates. In this work, we have selected a galaxy, NGC~3319, where we can apply a wealth of independent black hole mass estimates. NGC~3319 is a gas-rich, bulgeless, late-type galaxy. It is a strongly-barred spiral galaxy classified as SBcd(rs) \citep{RC3} and has its bar aligned with the major axis \citep{Randriamampandry:2015}. Moreover, \citet{Jiang:2018} identify it as possessing a low-luminosity AGN with a high-accretion-rate signalled by a nuclear X-ray point source and assume a black hole mass between $3\times10^2\,\mathrm{M_{\odot}}$ and $3\times10^5\,\mathrm{M_{\odot}}$ based on a high Eddington ratio of 1 to $10^{-3}$, despite a non-detection in the radio. Using the X-ray variability, they report an estimate of $\sim$10$^{5\pm2}$\,M$_{\odot}$, and using the `fundamental plane of black hole activity', they reported an upper limit of 10$^5$\,M$_{\odot}$  in the absence of radio data. NGC~3319 had previously been recognised as a possible low-ionisation nuclear emission-line region (LINER) galaxy \citep{Heckman:1980,Pogge:1989}, or at least it possessed an uncertain H\,\textsc{ii} nucleus \citep{Ho:1997}. Recently, \citet{Baldi:2018} classified its nuclear type as a LINER based on BPT \citep{BPT} diagram diagnostics. This classification is of significance since AGN with black holes are suspected sources of stimulating LINER spectral emission \citep{Heckman:1980b}.

In this study, we endeavour to constrain better the mass of the potential IMBH in the nucleus of NGC~3319 via a meta-analysis of multiple mass estimates based on independently measured quantities. In the numerous subsections of Section~\ref{sec:bhme}, we present a detailed analysis and application of ten separate black hole mass scaling relations and ultimately combine these estimates to yield an overall black hole mass estimate with confidence limits. The uncertainty on each mass estimate is used to weight every estimate before combining the results, via standard statistical techniques, to obtain the final mass estimate whose uncertainty is naturally less than that of the individual mass estimates. In the final section (Section~\ref{sec:disc}), we discuss the results of our investigation, comment on the implications, and remark on the benefit from continued study of NGC~3319.

Following \citet{Jiang:2018}, we adopt a redshift-independent luminosity distance of $14.3\pm1.1$\,Mpc \citep[Cepheid variable star distance from][]{Sakai:1999}, with a physical scale of $69\pm5$\,pc\,arcsec$^{-1}$. All values from the literature have been adjusted to accommodate our adopted distance to NGC~3319. Black hole masses ($M_\bullet$) and other masses throughout this work are represented as logarithmic (solar) mass values, i.e. $\mathcal{M}\equiv\log{M}$, where $M$ is mass in units of solar masses (M$_\odot$). All uncertainties are presented as (or have been scaled to) $1\,\sigma \approx 68.3\%$ confidence intervals. All magnitudes are given in the AB system \citep{Oke:1974}.

\section{Black Hole Mass Estimates}\label{sec:bhme}

In the following subsections (\ref{sec:m-phi}--\ref{sec:colour}), we applied ten different black hole mass scaling relations to estimate the mass of the black hole (NGC~3319*) residing at the centre of NGC~3319. We use the latest, and thus in some instances morphology-dependent, black hole scaling relations. Although the use of reverberation mapping has revealed that AGN extend the $M_\bullet$--$M_{\rm bulge,\star}$ relation to black hole masses of $10^5\,\mathrm{M_{\odot}}$ \citep{Graham:2015}, the paucity of confirmed IMBHs (and thus their dearth in the construction of black hole mass scaling relations) requires us to extrapolate these relations to reach into the IMBH regime.\footnote{This is also the case with reverberation mapping, which assumes the $f$-factor (used to convert virial products into virial masses) holds constant.} Albeit, we note that NGC~205 \citep{Nguyen:2019} and NGC~404 \citep{Nguyen:2017} now extend the relations down to $\sim$10$^4$ and $\sim$10$^5$\,M$_{\odot}$, respectively. In Section~\ref{sec:PDF}, we combine the black hole mass estimates, accounting for the different levels of scatter in each estimate.

\subsection{The $M_\bullet$--$\phi$ relation}\label{sec:m-phi}

The aesthetic beauty of `spiral nebulae' has been observed for 175 years, since Lord Rosse's observations of the Whirlpool Galaxy (NGC~5194). However, significant mysteries still abound between the nature of these striking features and properties of their host galaxies \citep{D'Onghia:2013}. The seminal works that established the spiral density wave theory \citep{Lin:1964,Lin:1966,Lin:1969} have provided perhaps the most lucid and lasting explanation of (grand design) spiral genesis. Indeed, the spiral density theory has been supported by observations in numerous studies \citep{Davis:2015,Pour-Imani:2016,Yu:2018,Peterken:2019,Miller:2019,Vallee:2019,Vallee:2020,Abdeen:2020,Griv:2020,Griv:2021}.

In particular, \citet{Lin:1966} predicted that the geometry of spiral patterns should be governed by two primary galactic properties: (i) the density of the galactic disk and (ii) the central gravitational potential (mass) of the galaxy. Specifically, the pitch angle of the spiral pattern at a distance $R$ from a galaxy's centre should be directly proportional to the density of the disk at $R$ and inversely proportional to the mass of the galaxy $\leq$$R$. \citet{Davis:2015} tested this prediction and found a tight trivariate relationship between the pitch angle, the stellar bulge mass, and the neutral atomic hydrogen density in the disk of a galaxy. Additional studies pertaining to dark matter halos have also shown a correlation between pitch angle and the central mass concentration, as determined by the shear of the rotation curve of a galaxy \citep{Seigar:2006,Seigar:2014}. These theoretical and observational studies provide perhaps the best explanations of why the pitch angle correlates with its host galaxy: the pitch angle is clearly related to the central mass of a galaxy, of which the `barge' (bar and bulge) and black hole are integral components entwined via coevolution.

\begin{figure*}
\includegraphics[width=\textwidth]{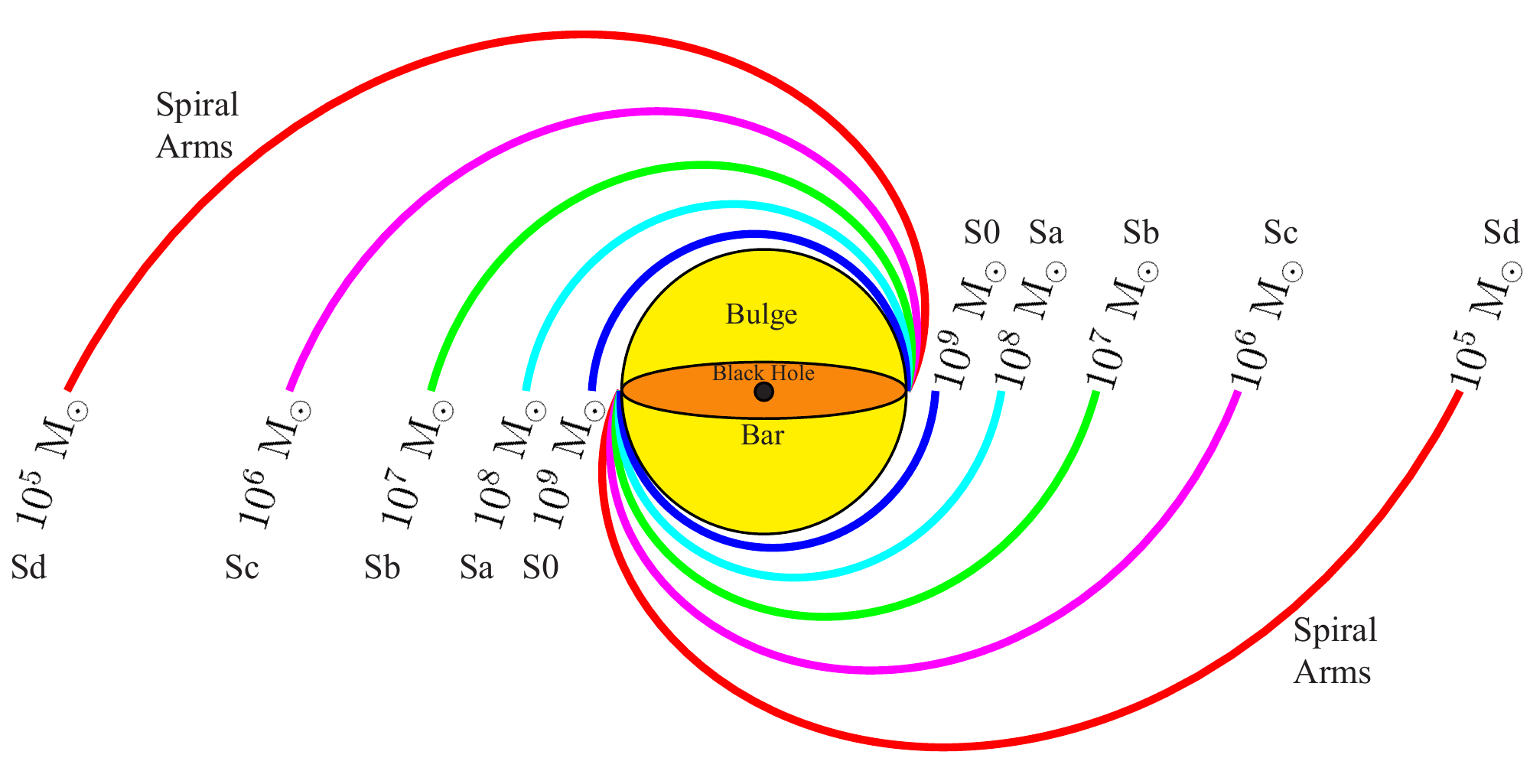}
\caption{Spiral galaxy arms with varying degrees of tightness, with the corresponding galaxy morphological type and central black hole mass in units of our Sun's mass. This template can be used to estimate central black hole masses in spiral galaxies. The outermost spiral (\textcolor{red}{{\hdashrule[0.35ex]{6mm}{0.4mm}{}}}) has $|\phi|=26\fdg7$, which is indicative of a central black hole with a mass of $10^5\,\mathrm{M_{\odot}}$ via equation~8 from \citet{Davis:2017}.}
\label{fig:M-phi}
\end{figure*}

The geometry of logarithmic spirals closely matches the shape of spiral arms in galaxies. Quantitatively, the shape (tightness of winding) of a logarithmic spiral is governed by the absolute value of its pitch angle,\footnote{For introductory reading on pitch angle, see section~2 from \citet{Davis:2017}.} $|\phi|$, as introduced by \citet{Pahlen:1911}. \citet{Seigar:2008} first presented evidence of a strong relationship between pitch angle and the mass of a spiral galaxy's central black hole. As the sample of spiral galaxies with directly-measured black hole masses grew incrementally in size over the years, \citet{Berrier:2013} and later \citet{Davis:2017} presented refinements to the $M_\bullet$--$\phi$ relation. A graphical representation of the relation found by \citet[][equation~8]{Davis:2017} is shown in Fig.~\ref{fig:M-phi}.
We employ equation~8 from \citet{Davis:2017} to convert measured pitch angles into black hole masses, including an intrinsic scatter of 0.33\,dex (added in quadrature with a full propagation of errors on the pitch angle measurement, as well as errors on the slope and intercept of the relation).

The existence of an $M_\bullet$--$\phi$ relation has been seen not only in observations \citep{Seigar:2008,Berrier:2013,Davis:2017} but also in simulations. \citet{Burcin:2018} measured the pitch angles for a random sample of 95 galaxies drawn from the \textit{Illustris} simulation \citep{Vogelsberger:2014} and recovered an $M_\bullet$--$\phi$ relation that was consistent with that found from observational studies. Thus, the nascent $M_\bullet$--$\phi$ relation has already garnered empirical and theoretical (via theory and simulations) support to become a full-fledged black hole mass scaling relation. Its progress has proliferated in only a dozen years; future improvements in observations and sample size should add to its established legitimacy. The search for the primary relation with black hole mass continues, and the lack of a spiral pattern in early-type galaxies rules out the $M_\bullet$--$\phi$ relation, just as the absence of bulges in some late-type galaxies negates the $M_\bullet$--$M_{\rm bulge,\star}$ relation. Nonetheless, the low level of scatter in both relations make them valuable black hole mass estimators.

Several software programs have been devised to handle the quantitative measurement of spiral galaxy pitch angle. In this work, we utilise three of the most prominent and robust packages to measure pitch angle: \textsc{2dfft} \citep{Davis:2012,2dfft,p2dfft}, \textsc{spirality} \citep{Shields:2015,spirality}, and \textsc{sparcfire} \citep{sparcfire}. Each code uses an independent method of measuring pitch angle, each with its unique advantage.\footnote{For a demonstrative comparison of each software package, see the appendix from \citet{Davis:2017}.} Each routine measures pitch angle after the original galaxy image (Fig.~\ref{fig:sparcfire}, left panel) has been deprojected to an artificial \emph{face-on} orientation (Fig.~\ref{fig:sparcfire}, middle panel). We adopt the outer isophote position angle ($PA_{\rm outer}$, degrees east of north) and ellipticity ($\epsilon_{\rm outer}$) values for NGC~3319 from \citet{Salo:2015}: $PA_{\rm outer}=43\fdg0\pm0\fdg7$ and $\epsilon_{\rm outer}=0.435\pm0.003$. This ellipticity is equivalent to an inclination of the disk, $i_{\rm disk} \equiv \cos^{-1}(1-\epsilon_{\rm outer})=55\fdg6\pm0\fdg2$.

We measured the pitch angles from a \textit{Spitzer Space Telescope} Infrared Array Camera (IRAC) $8.0\,\micron$ image obtained from the \textit{Spitzer Heritage Archive}.\footnote{\url{https://irsa.ipac.caltech.edu/applications/Spitzer/SHA}} Recent studies \citep{Pour-Imani:2016,Miller:2019} have presented observational evidence that 8.0-$\micron$ light highlights the physical location of the spiral density wave in spiral galaxies. 8.0-$\micron$ light comes from the glow of warm dust around nascent natal star-forming regions that have been shocked into existence by the spiral density wave.

\subsubsection{\textsc{sparcfire}}\label{sec:sparcfire}

\textsc{sparcfire} \citep{sparcfire} uses computer vision techniques to identify the pixel clusters that form the architecture of spiral arms in spiral galaxies and fits logarithmic spiral segments to the clusters. \textsc{sparcfire} classifies each spiral based on its chirality: Z-wise, spirals that grow radially in a counterclockwise direction ($\phi<0$); and S-wise, spirals that grow radially in a clockwise direction ($\phi>0$). Based on the number and arc lengths of the ensemble of fitted spirals, we adopted a dominant chirality for the galaxy and ignored all spurious arcs matching the secondary chirality. We calculated a weighted-arithmetic-mean pitch angle for the galaxy based on a weight for each arc ($w_i$) such that $w_i\equiv s_i/r_{0,i}$, where $s_i$ is the arc length and $r_{0,i}$ is the inner radius (from the origin at the galactic centre) for an individual arc segment. Therefore, the highest weighting resides with long arcs near the centre of the galaxy and short, possibly spurious arc segments in the outer region of the galaxy, are made insignificant.

\begin{figure*}
\includegraphics[clip=true, trim= 1.25mm 0.45mm 1.0mm 1.8mm, width=0.328\textwidth]{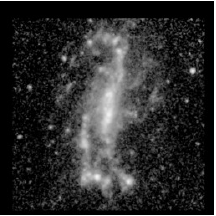}
\includegraphics[clip=true, trim= 0.8mm 0mm 0.5mm 1.4mm, width=0.33\textwidth]{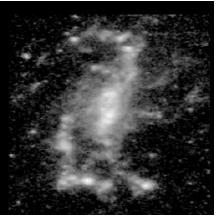}
\includegraphics[clip=true, trim= 0.8mm 0mm 0.5mm 1.4mm, width=0.33\textwidth]{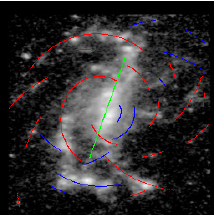}
\caption{{\bf Left (Original)} -- \textit{Spitzer} $8.0\,\micron$ image of NGC~3319. Here, the image has been aligned, pointing the top of the image in the direction of the galaxy's position angle ($43\fdg0$ east of north), and the image has been cropped into a square that is $5\arcmin \times 5\arcmin$ ($20.7\,{\rm kpc} \times 20.7\,{\rm kpc}$). {\bf Middle (Deprojected)} -- here, the original image has been deprojected to an artificial \emph{face-on} orientation, achieved by stretching the X-axis by a factor of $a/b\equiv(1-\epsilon_{\rm outer})^{-1}=1.77$, where $a$ is the semi-major axis length, and $b$ is the semi-minor axis length of the outer isophotes \citep{Salo:2015}. {\bf Right (Spiral Arcs)} -- the spiral arcs measured by \textsc{sparcfire} \citep{sparcfire} are overlaid upon the deprojected image. Fitted lines depict: (used) Z-wise spiral arcs (\textcolor{red}{{\hdashrule[0.35ex]{6mm}{0.4mm}{}}}), (ignored) S-wise spiral arcs (\textcolor{blue}{{\hdashrule[0.35ex]{6mm}{0.4mm}{}}}), and the galactic bar (\textcolor{green}{{\hdashrule[0.35ex]{6mm}{0.4mm}{}}}). The reported pitch angle, $31\fdg7\pm4\fdg5$, is the weighted-mean pitch angle of the dominant-chirality \textcolor{red}{red} spiral arcs (see Section~\ref{sec:sparcfire}).}
\label{fig:sparcfire}
\end{figure*}

As seen in the right panel of Fig.~\ref{fig:sparcfire}, the dominant chirality is Z-wise.
We computed the pitch angle and converted it to a black hole mass prediction via the $M_\bullet$--$\phi$ relation as follows,
\begin{IEEEeqnarray}{rCl}
|\phi|_\textsc{sparcfire}&=&31\fdg7\pm4\fdg5\rightarrow \nonumber \\
\mathcal{M}_\bullet(|\phi|_{\textsc{sparcfire}})&=&4.15\pm0.86.
\label{eqn:sparcfire}
\end{IEEEeqnarray}

\subsubsection{\textsc{2dfft}}

\textsc{2dfft} \citep{Davis:2012,2dfft,p2dfft} is a two-dimensional fast \textit{Fourier} transform software package that decomposes a galaxy image into logarithmic spirals. It computes the amplitude of each \textit{Fourier} component by decomposing the observed distribution of light in an image into a superposition of logarithmic spirals as a function of pitch angle, $\phi$, and harmonic-mode, $m$, i.e. the order of rotational symmetry (e.g. two-fold, three-fold, and higher-order symmetries). For the \emph{face-on} view of NGC~3319 (Fig.~\ref{fig:sparcfire}, middle panel), the maximum amplitude is achieved with $m=2$ (i.e. two spiral arms) and
\begin{equation}
|\phi|_\textsc{2dfft}=28\fdg4\pm3\fdg5\rightarrow \mathcal{M}_\bullet(|\phi|_\textsc{2dfft})=4.72\pm0.70.
\label{eqn:2dfft}
\end{equation}

\subsubsection{\textsc{spirality}}

\textsc{spirality} \citep{Shields:2015,spirality} is a template-fitting software. Given a \emph{face-on} image of a spiral galaxy, it computes a library of spiral coordinate systems with varying pitch angles. For NGC~3319 (Fig.~\ref{fig:sparcfire}, middle panel), the best-fitting spiral coordinate system has a pitch angle of
\begin{IEEEeqnarray}{rCl}
|\phi|_\textsc{spirality}&=&24\fdg4\pm4\fdg1\rightarrow \nonumber \\
\mathcal{M}_\bullet(|\phi|_{\textsc{spirality}})&=&5.40\pm0.74.
\label{eqn:spirality}
\end{IEEEeqnarray}

\subsubsection{Weighted-mean pitch angle and black hole mass}\label{sec:wm_phi}

The $M_\bullet$--$\phi$ relation is a tight relation, with intrinsically low scatter. However, the slope of the relation is relatively steep, and thus small changes in pitch angle equate to large changes in black hole mass. Specifically, a change in pitch angle of only $5\fdg8$ is associated with a 1.0\,dex change in black hole mass. For late-type spiral galaxies like NGC~3319, their open spiral structures often feature inherent flocculence and asymmetries amongst individual spiral arms. Furthermore, due to the diminished total masses of these galaxies (as compared to early-type spiral galaxies), galaxy harassment and tidal interactions are more impactful in disrupting their spiral structures.

The average uncertainty amongst our equations~\ref{eqn:sparcfire}--\ref{eqn:spirality} is $4\fdg0$ (a difference of 0.68\,dex in black hole mass). Nonetheless, all three of the pitch angle measurements possess overlapping error bars. To produce a more robust pitch angle measurement, we combine all three measurements (equations~\ref{eqn:sparcfire}--\ref{eqn:spirality}) to yield a weighted-arithmetic-mean pitch angle,
\begin{equation}
\bar{\phi}=\frac{\sum_{i=1}^{N} w_{i} \phi_i}{\sum_{i=1}^{N} w_{i}},
\label{eqn:wm_phi}
\end{equation}
and associated uncertainty,
\begin{equation}
\delta\bar{\phi}=\frac{\sqrt{\sum_{i=1}^{N}(w_i\delta \phi_i)^2}}{\sum_{i=1}^{N}w_i}=\sqrt{\frac{1}{\sum_{i=1}^{N}w_i}},
\label{eqn:wm_phi_error}
\end{equation}
with a weight for each measurement that is inversely proportional to the square of the uncertainty of its measurement, i.e. inverse-variance weighting, $w_i=\left(\delta\phi_i\right)^{-2}$. This yields
\begin{equation}
|\bar{\phi}|=28\fdg0\pm2\fdg3\rightarrow \mathcal{M}_\bullet(|\bar{\phi}|)=4.79\pm0.54.
\label{eqn:phi_wm}
\end{equation}

Our use of the independent black hole mass scaling relations, and their reported $\pm1\,\sigma$ scatter, assumes a normal distribution for each. Assuming a normal distribution for our weighted-mean, we can then calculate the probability of having an IMBH. Given a mass estimate for a black hole and its associated error ($\delta\mathcal{M}_\bullet$), we can compute the probability that the black hole is less-than-supermassive ($\mathcal{M}_\bullet\leq5$) as follows,
\begin{equation}
P(\mathcal{M}_\bullet\leq5)=\frac{1}{2}\left[1+{\rm erf}\left(\frac{5-\mathcal{M}_\bullet}{\delta\mathcal{M}_\bullet\sqrt{2}}\right)\right]
\label{eqn:prob}
\end{equation}
\citep{Weisstein:2002}. Doing so for the mass estimate from equation~(\ref{eqn:phi_wm}), we find $P(\mathcal{M}_\bullet\leq5)=65\%$. We have additionally checked the pitch angle in alternative imaging that also traces star formation in spiral arms, by using the Galaxy Evolution Explorer (GALEX) far-ultraviolet (FUV) passband (1350--1750\,\AA). We found that the pitch angle from GALEX FUV imaging, $27\fdg5\pm3\fdg9$, is highly consistent with that from 8.0-$\micron$ imaging.

\subsection{The $M_\bullet$--$M_{{\rm gal},\star}$ relation}

For our second estimate, we used the total stellar mass of NGC~3319 as a predictor of the black hole mass at its centre. We began by obtaining \textit{Spitzer} images and masks for NGC~3319 from the S$^4$G catalogue \citep{Sheth:2010}.\footnote{\url{https://irsa.ipac.caltech.edu/data/SPITZER/S4G/index.html}} We elected to use the $3.6\,\micron,\star$ stellar image from \cite{Querejeta:2015}. The $3.6\,\micron,\star$ image has been created after determining the amount of glowing dust present (by analysing the empirical $3.6\,\micron$ and $4.5\,\micron$ images) and subsequently subtracting the dust light from the $3.6\,\micron$ image. Thus, the $3.6\,\micron,\star$ image shows only the light emitted from the stellar population, and its luminosity can be directly converted into a stellar mass. We adopted a $3.6\,\micron$ stellar mass-to-light ratio, $\Upsilon_{3.6\,\micron,\star}=0.60\pm0.09$ from \citet{Meidt:2014},\footnote{The $3.6\,\micron$ bandpass has a low uncertainty for the stellar mass-to-light ratio, with $\Upsilon_\star$ from 0.40 to 0.55 \citep{Schombert:2019}. This is consistent with the observed (i.e. with dust glow) $\Upsilon_{3.6\,\micron,\star,{\rm obs}}=0.453\pm0.072$ value derived by \citet[][section~2.8]{Davis:2019}, which is equivalent to the dust-corrected $\Upsilon_{3.6\,\micron,\star}=0.60\pm0.09$ from \citet{Meidt:2014}.} along with a solar absolute magnitude, $\mathfrak{M}_{3.6\,\micron,\odot}=6.02$\,mag (AB), at $3.6\,\micron$ \citep{Oh:2008}.

\begin{figure*}
\includegraphics[clip=true, trim= 0mm 0mm 0mm 0mm, width=0.246\textwidth]{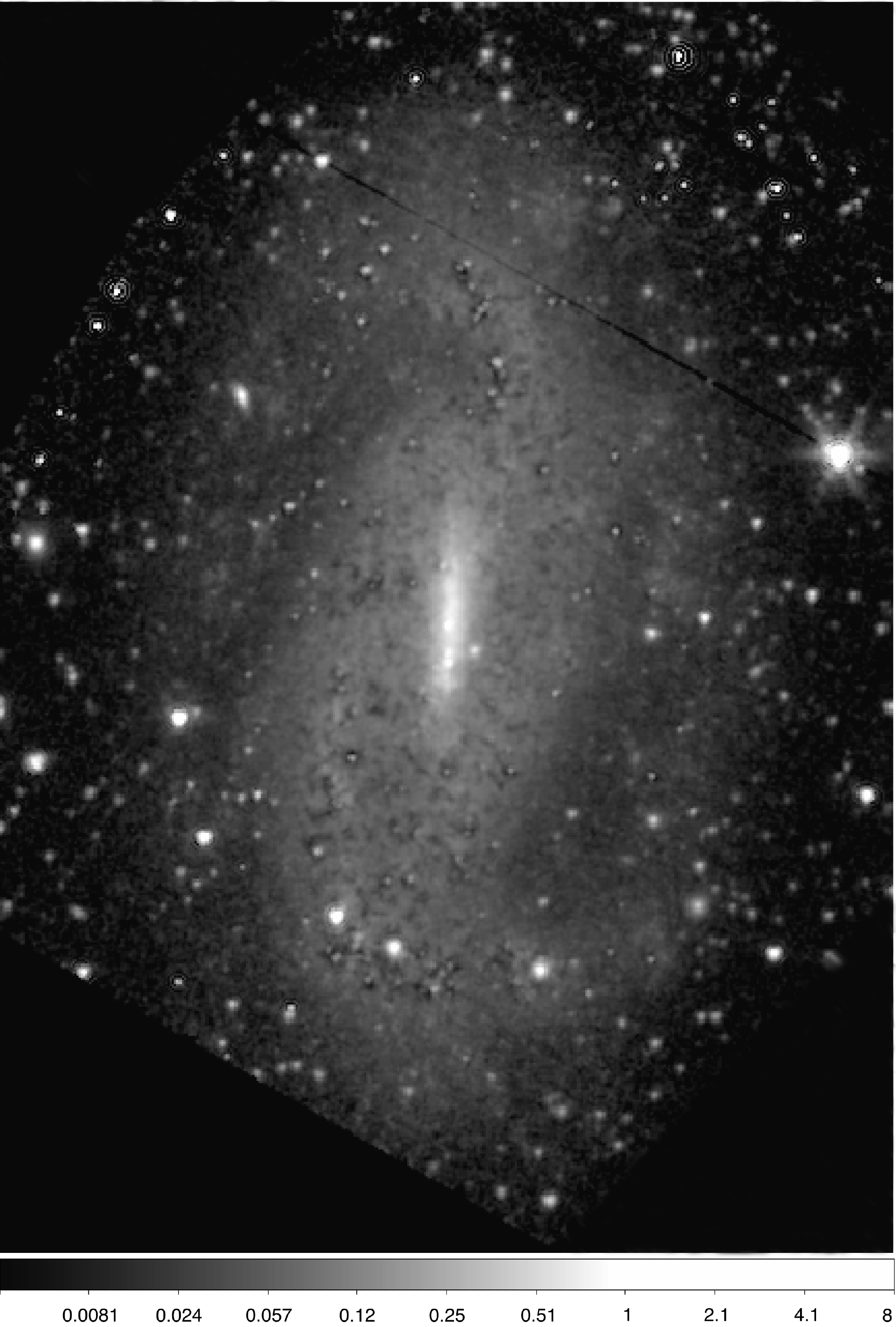}
\includegraphics[clip=true, trim= 0mm 0mm 0mm 0mm, width=0.246\textwidth]{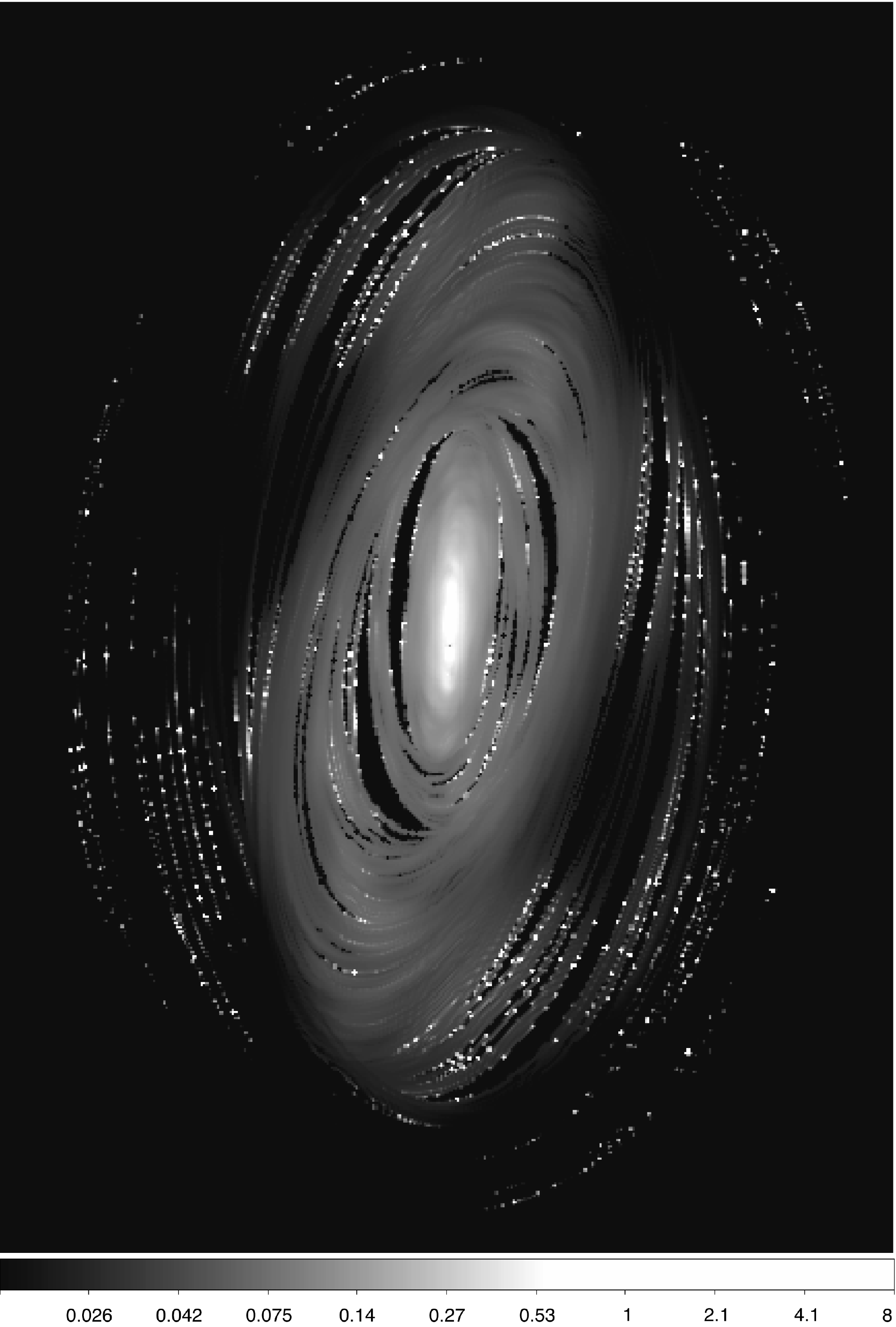}
\includegraphics[clip=true, trim= 0mm 0mm 0mm 0mm, width=0.246\textwidth]{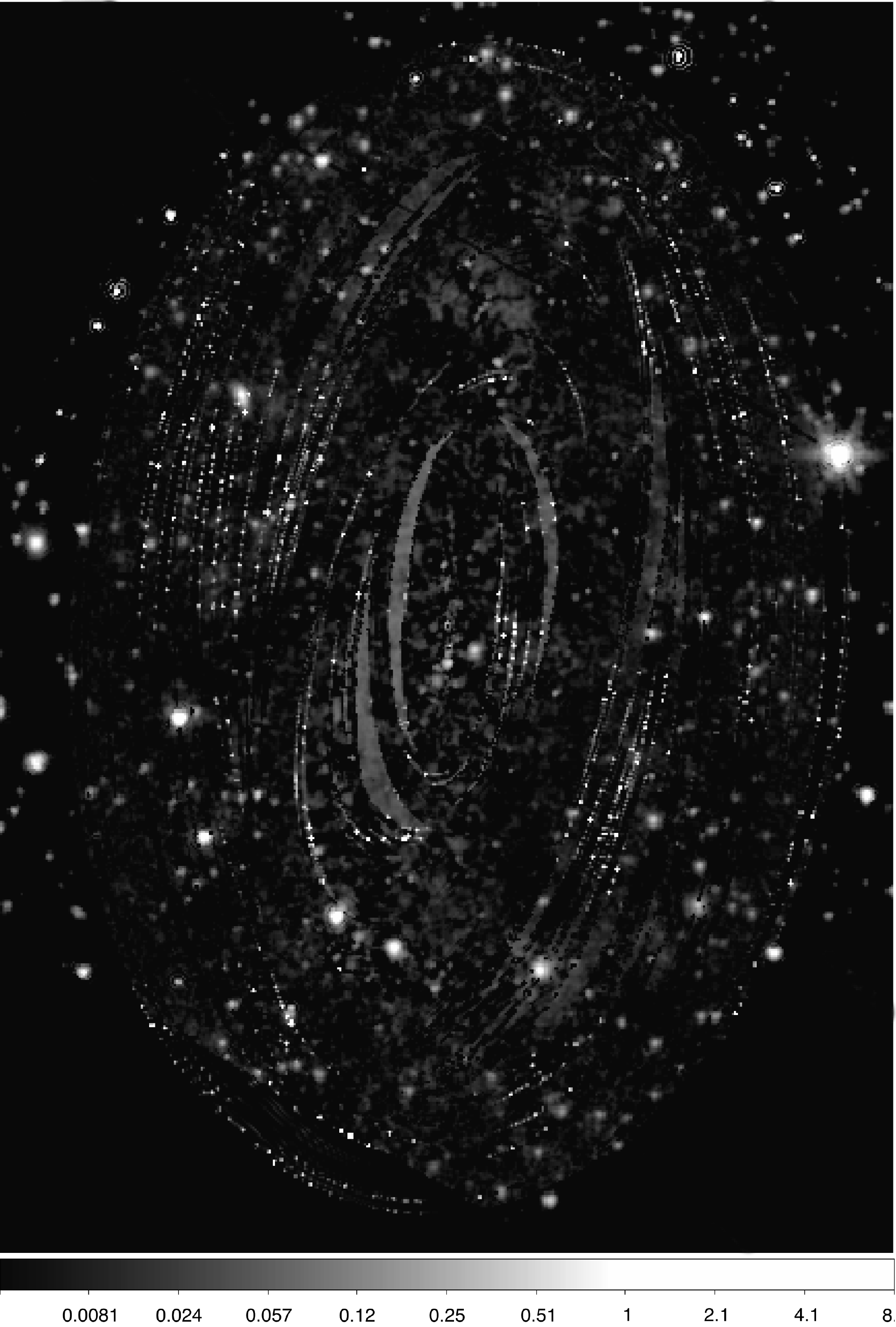}
\includegraphics[clip=true, trim= 0mm 0mm 0mm 0mm, width=0.246\textwidth]{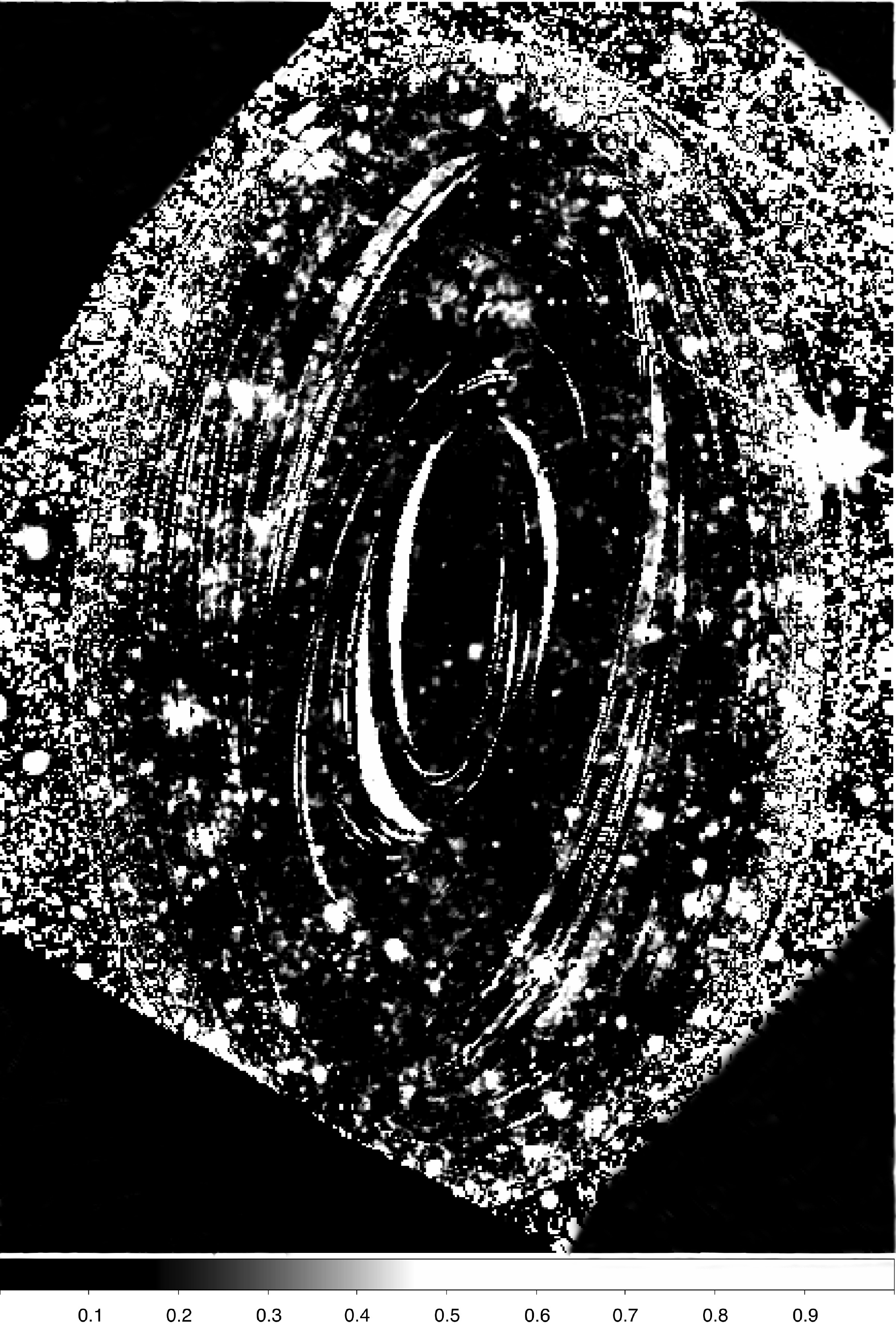}
\caption{{\bf Left (Original)} -- \textit{Spitzer} $3.6\,\micron,\star$ image of NGC~3319. Here, the image has been aligned so that the top of the image is pointing in the direction of the galaxy's position angle ($43\fdg0$ east of north), and the image has been cropped, so it is $5\arcmin \times 7\arcmin$ ($20.7\,{\rm kpc} \times 28.98\,{\rm kpc}$). The black pixels indicate no intensity, and white pixels (pixel intensity of 8.2\,MJy\,sr$^{-1}$) indicate $\mu_{3.6\,\micron,\star}\leq18.188$\,mag\,arcsec$^{-2}$. {\bf Second from Left (Model)} -- model produced by \textsc{isofit} and \textsc{cmodel} \citep{Ciambur:2015}, which includes a sky background of 0.0180\,MJy\,sr$^{-1}$ \citep{Salo:2015}. {\bf Second from Right (Residual)} -- residual image, such that Residual $\equiv$ Original $-$ Model. {\bf Right (Division)} -- division image, such that Division $\equiv$ Residual $\div$ Original. The division image depicts the relative difference between the original and the residual image. Pixel values are between zero (black) and one (white), representing maximal and minimal change, respectively.}
\label{fig:images}
\end{figure*}

To model the light from NGC~3319, we utilised the isophotal fitting and modelling software routines \textsc{isofit} and \textsc{cmodel} \citep{Ciambur:2015}, respectively. After masking extraneous light sources, we ran \textsc{isofit} on the $3.6\,\micron,\star$ image (Fig.~\ref{fig:images}, left panel) and used \textsc{cmodel} to extract, and create a representation of, the galaxy (Fig.~\ref{fig:images}, second panel). The quality of the extraction can be seen in the residual images presented in the right two panels of Fig.~\ref{fig:images}.

\begin{figure*}
\includegraphics[clip=true, trim= 10mm 3mm 3mm 5mm, width=0.246\textwidth]{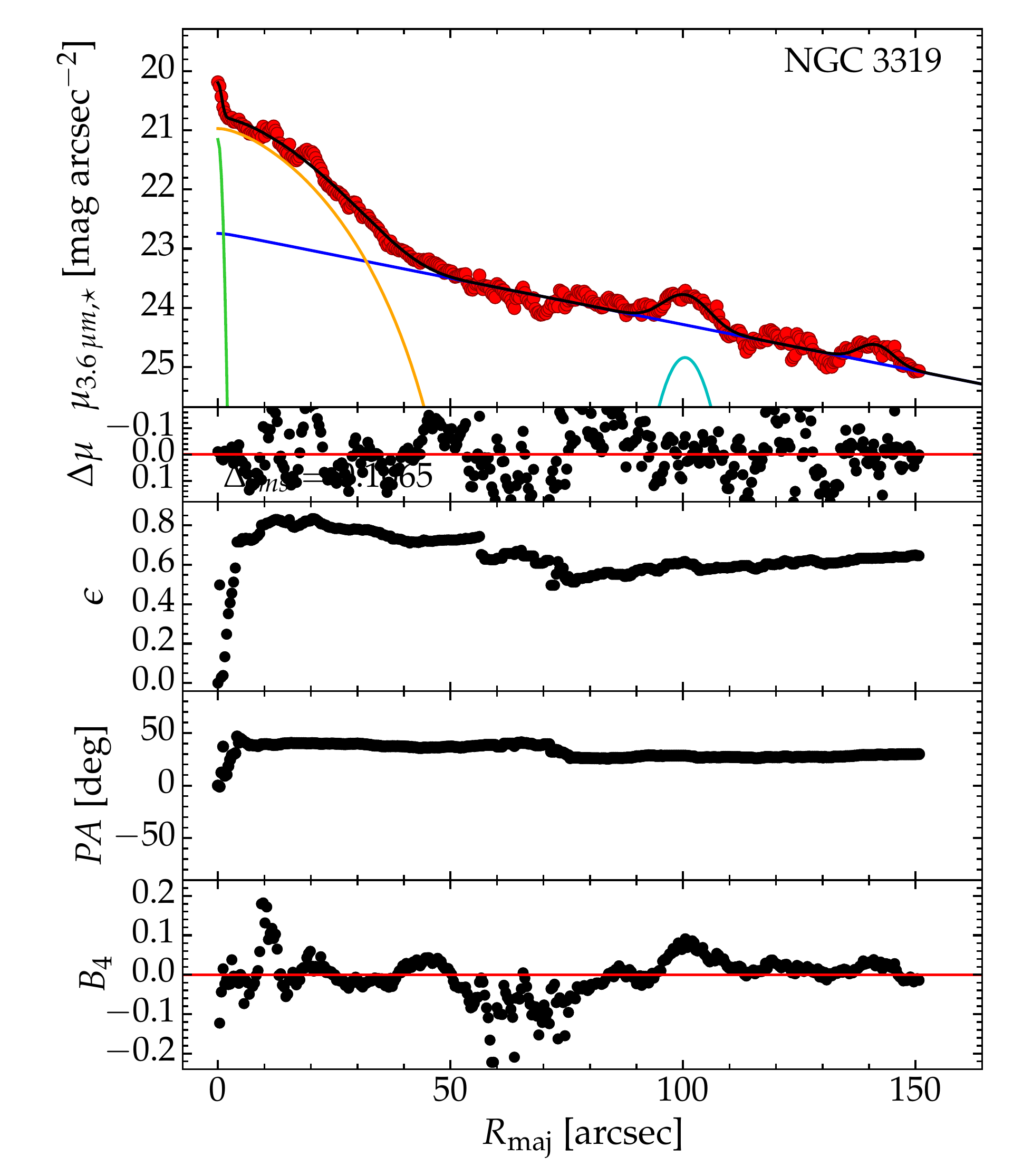}
\includegraphics[clip=true, trim= 10mm 3mm 3mm 5mm, width=0.246\textwidth]{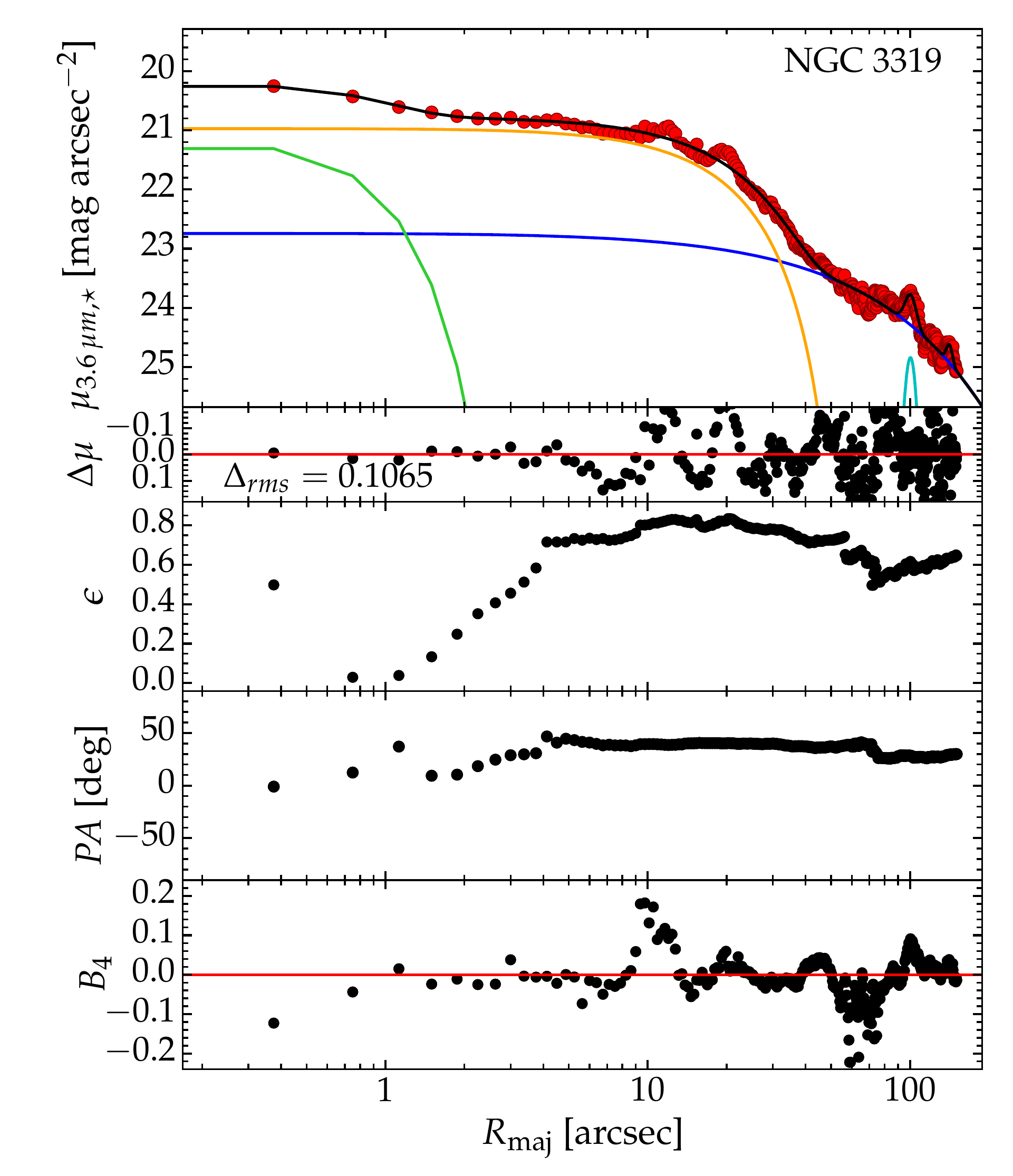}
\includegraphics[clip=true, trim= 10mm 3mm 3mm 5mm, width=0.246\textwidth]{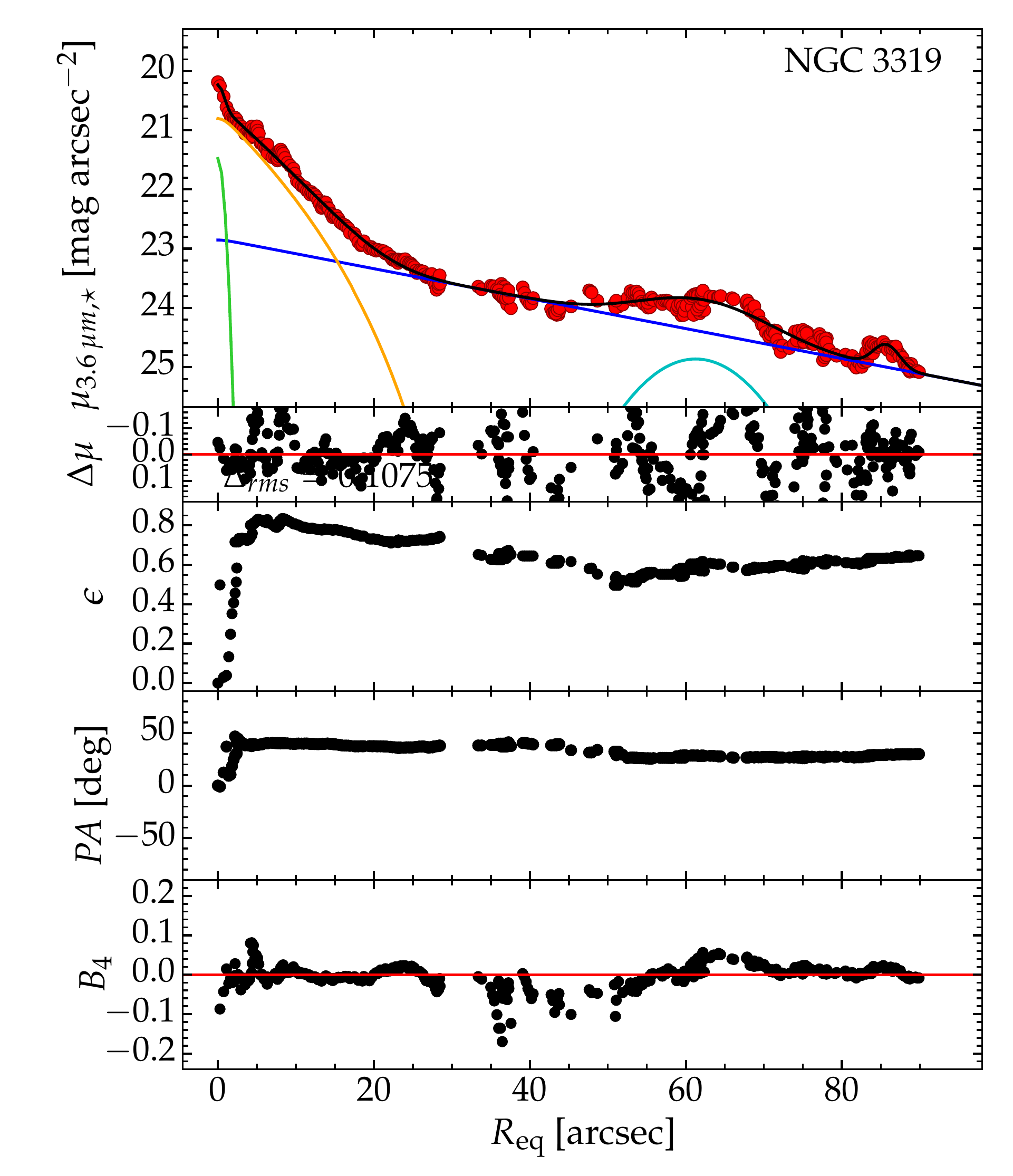}
\includegraphics[clip=true, trim= 10mm 3mm 3mm 5mm, width=0.246\textwidth]{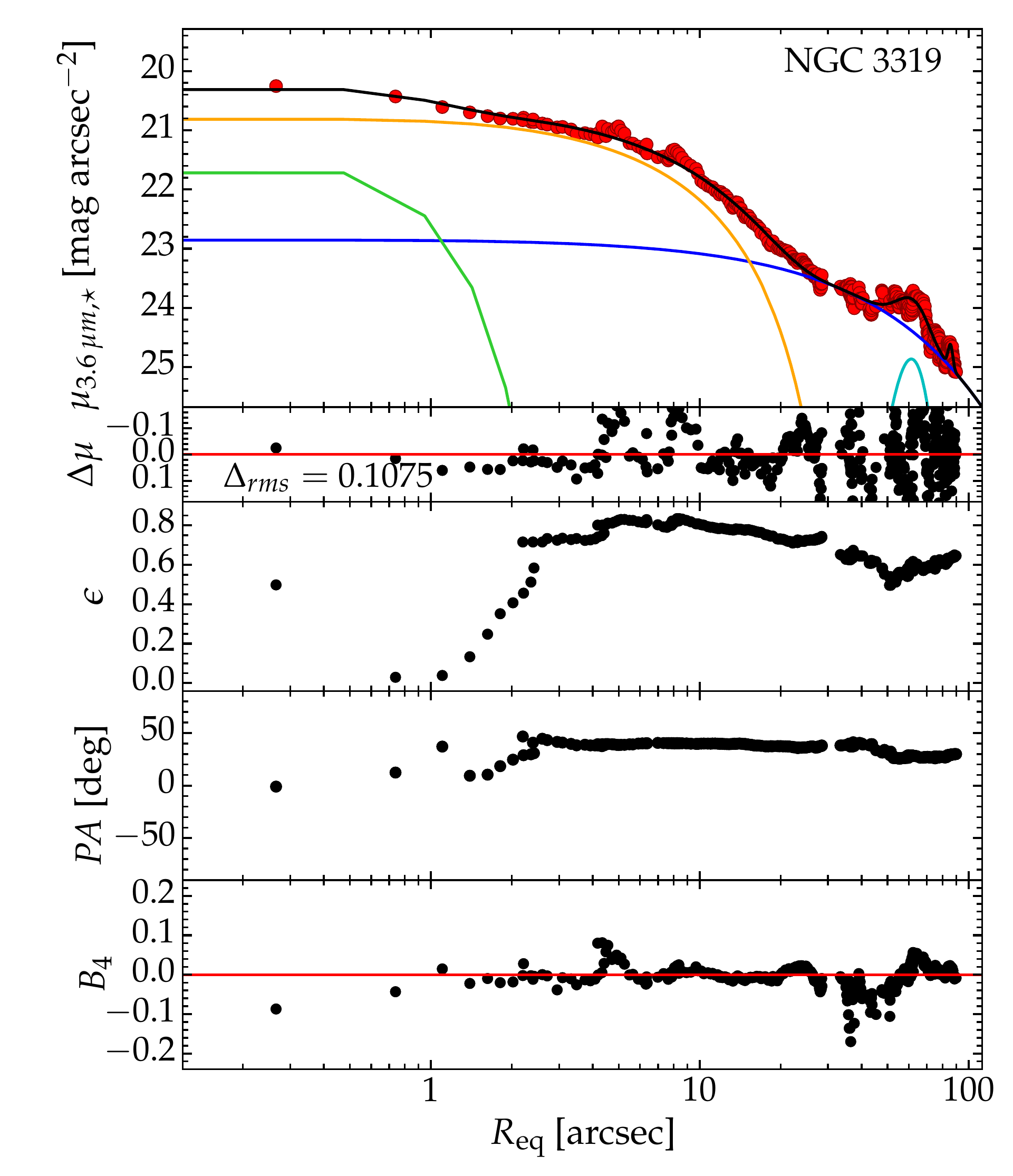}
\caption{Surface brightness profile decompositions produced by \textsc{profiler} \citep{Ciambur:2016}. {\bf Panels (from left to right):} linear major-axis, log major-axis, linear equivalent-axis, and log equivalent-axis profiles; $R_{\rm eq}=\sqrt{ab}= R_{\rm maj}\sqrt{1-\epsilon}$ and $R_{\rm maj}\equiv a$. {\bf Subplots (from top to bottom):} surface brightness profile (\textcolor{red}{$\bullet$}) and model (\hdashrule[0.35ex]{6mm}{0.5mm}{}) built from the summation of the following components: PSF (\textcolor{LimeGreen}{\hdashrule[0.35ex]{6mm}{0.4mm}{}}), bar (\textcolor{Orange}{\hdashrule[0.35ex]{6mm}{0.4mm}{}}), disk (\textcolor{blue}{\hdashrule[0.35ex]{6mm}{0.4mm}{}}), and spiral arms (\textcolor{cyan}{\hdashrule[0.35ex]{6mm}{0.4mm}{}}), the faint outer spiral arm (at $R_{\rm maj}\approx140\arcsec\equiv R_{\rm eq}\approx85\arcsec$) lies below the plotted region; residual profile with total rms scatter ($\Delta_{rms}$); ellipticity profile; position angle profile; and fourth-order cosine \textit{Fourier} harmonic coefficient, $B_4$ ($B_2$, $B_3$, $B_6$, $B_8$, and $B_{10}$ harmonics are also fit and contribute to the model).}
\label{fig:SBPs}
\end{figure*}

The extracted galaxy was then analysed by the surface brightness profile fitting software \textsc{profiler} \citep{Ciambur:2016}. This works by convolving the galaxy model with the \textit{Spitzer} (IRAC channel 1) point spread function (PSF) with a full width at half maximum (FWHM) of $1\farcs66$ for the cryogenic mission\footnote{\url{https://irsa.ipac.caltech.edu/data/SPITZER/docs/irac/iracinstrumenthandbook/5/}} until an optimal match is achieved.\footnote{\textsc{profiler} uses an unweighted least-squares Levenberg-Marquardt \citep{Marquardt:1963} algorithm \citep[via \textsc{python} package \textsc{lmfit};][]{Newville:2016} to minimise the total rms scatter, $\Delta_{rms}=\sqrt{(n-f+1)^{-1}\sum_{i=R_{\rm min}}^{R_{\rm max}}\left(\mu_{{\rm data},i}-\mu_{{\rm model},i}\right)^2}$, with surface brightnesses of the data (obtained from \textsc{isofit}) and model, each at radial bin, $i$, where $n$ is the number of radial bins (inclusive) between the minimum ($R_{\rm min}$) and maximum ($R_{\rm max}$) user-selected radii, and $f$ is the number of free parameters (i.e. the number of user-selected components); \textsc{profiler} adjusts the model (summation of user-selected components) until a global minimum is reached. Additionally, a residual profile, $\Delta\mu(R) = \mu_{\rm data}(R)-\mu_{\rm model}(R)$, is provided in the output plots of \textsc{profiler} to demonstrate the quality of the fit as a function of galactocentric radius ($R$).} We present the resulting galaxy surface brightness profiles and multi-component fits for both the major axis (Fig.~\ref{fig:SBPs}, left two panels) and the geometric mean axis, equivalent to a circularised representation of the galaxy (Fig.~\ref{fig:SBPs}, right two panels).

We confirm that NGC~3319 is a bulgeless galaxy and does not require a traditional S\'ersic bulge component \citep{Sersic:1963,Ciotti:1991,Graham:2005}. Instead, we generate a convincing fit that adequately captures all of the light of the galaxy (with a total rms scatter, $\Delta_{rms}<0.11$\,mag) using five components: a \textit{Ferrers} bar \citep{Ferrers:1877}; an exponential disk; two Gaussian components to capture spiral arm crossings of the major axis; and a point source at the centre. We calculate a total integrated $3.6\,\micron,\star$ apparent magnitude of $12.42\pm0.11$\,mag (AB). Additional component magnitudes are tabulated in Table~\ref{tab:mags}. Based on its distance ($14.3\pm1.1$\,Mpc), we determine an absolute magnitude of $-18.37\pm0.20$\,mag for the galaxy at $3.6\,\micron,\star$. Applying $\Upsilon_{3.6\,\micron,\star}=0.60\pm0.09$ \citep{Meidt:2014} and $\mathfrak{M}_{3.6\,\micron,\odot}=6.02$\,mag (AB) yields a total logarithmic stellar mass of $\mathcal{M}_{{\rm gal},\star}=9.53\pm0.10$ \citep[cf.][$\mathcal{M}_{{\rm gal},\star}=9.53\pm0.16$]{Georgiev:2016} for NGC~3319.

\begin{table*}
\centering
\caption{NGC~3319 component magnitudes and masses.
\textbf{Columns:}
{\bf (1)} Surface brightness profile component.
{\bf (2)} $3.6\,\micron,\star$ apparent magnitude (AB).
{\bf (3)} $3.6\,\micron,\star$ absolute magnitude (AB).
{\bf (4)} Logarithmic (solar) mass.
}
\label{tab:mags}
\begin{tabular}{lccc}
\hline
Component & $\mathfrak{m}_{3.6\,\micron,\star}$ & $\mathfrak{M}_{3.6\,\micron,\star}$ & $\mathcal{M}_\star$ \\
 & (mag) & (mag) & (dex) \\
(1) & (2) & (3) & (4) \\
\hline
PSF (NC) & $20.22\pm0.32$ & $-10.57\pm0.36$ & $6.41\pm0.16$ \\
Bar & $14.70\pm0.53$ & $-16.09\pm0.56$ & $8.62\pm0.23$ \\
Disk & $12.67\pm0.10$ & $-18.12\pm0.20$ & $9.43\pm0.10$ \\
Inner Spiral & $15.20\pm0.10$ & $-15.59\pm0.20$ & $8.42\pm0.10$ \\
Outer Spiral & $17.63\pm0.29$ & $-13.16\pm0.34$ & $7.45\pm0.05$ \\
\textbf{Total} & \bm{$12.42\pm0.11$} & \bm{$-18.37\pm0.20$} & \bm{$9.53\pm0.10$} \\
\hline
\end{tabular}
\end{table*}

\citet{Savorgnan:2016} discovered a distinct red and blue sequence for early- and late-type galaxies in the $M_\bullet$--$M_{{\rm gal},\star}$ diagram, forming a revision to the core-S\'ersic (giant early-type galaxies) and S\'ersic (spiral and low-mass early-type galaxy) sequence from \citet{Graham:2012}, \citet{Graham:2013}, and \citet{Scott:2013}. \citet{Bosch:2016} subsequently showed this separation including additional galaxies, albeit with less reliable black hole masses, while \citet{Terrazas:2016} captured it in terms of star formation rate. Here, we apply the latest relation established for spiral galaxies with directly-measured black hole masses. Applying equation~3 (with $\upsilon\equiv1$) from \citet{Davis:2018}, this total galaxy stellar mass predicts a central black hole mass as follows,
\begin{equation}
\mathcal{M}_{{\rm gal},\star}=9.53\pm0.10\rightarrow \mathcal{M}_\bullet(M_{{\rm gal},\star})=3.38\pm1.02,
\label{eqn:Mgal}
\end{equation}
with $P(\mathcal{M}_\bullet\leq5)=94\%$.

As can be seen in the images and from the ellipticity profile, there is no mistaking that NGC~3319 possesses a strong bar that accounts for most of the light from the inner $R_{\rm maj}\lesssim30\arcsec$ ($\lesssim$2.1\,kpc) region of the galaxy. There is no obvious evidence of a bulge (spheroid) component; thus, NGC~3319 is considered to be a bulgeless galaxy. Even if one were to describe the bar as a pseudobulge mistakenly, its logarithmic `bulge' mass would only be $\mathcal{M}_{\rm bulge,\star}=8.62\pm0.23$ (see Table~\ref{tab:mags}). If applied to the $M_\bullet$--$M_{\rm bulge,\star}$ relation from \citet[][their equation~11]{Davis:2019}, this would still comfortably predict an IMBH of $\mathcal{M}_\bullet=3.73\pm0.91$, with {$P(\mathcal{M}_\bullet\leq5)=92\%$.

\subsection{The $M_\bullet$--$M_{{\rm NC},\star}$ relation}\label{sec:NC}

From our surface brightness profile decomposition of NGC~3319, we extracted a central point source apparent magnitude of $\mathfrak{m}_{3.6\,\micron,\star}=20.22\pm0.32$, yielding an absolute magnitude of $\mathfrak{M}_{3.6\,\micron,\star}=-10.57\pm0.36$. We will assume that this luminosity is due to the nuclear cluster (NC) of stars. Of course, some contribution of flux will come from the AGN. Therefore, we estimate an upper limit to the nuclear star cluster mass using $\Upsilon_{3.6\,\micron,\star}=0.60\pm0.09$ and $\mathfrak{M}_{3.6\,\micron,\odot}=6.02$\,mag (AB), to give $\mathcal{M}_{\rm NC,\star}\leq6.41\pm0.16$. We deem this to be a reasonable estimate since it lies between the recent estimates of $\mathcal{M}_{\rm NC,\star}=6.24\pm0.07$ \citep{Georgiev:2014,Georgiev:2016} and $\mathcal{M}_{\rm NC,\star}=6.76\pm0.07$ \citep{Jiang:2018}, both from \textit{Hubble Space Telescope} imaging of NGC~3319.

Using the new $M_\bullet$--$M_{{\rm NC},\star}$ relation of \citet{Graham:2019d},\footnote{See also \citet{Graham:2016} and equation~12 from \citet{Graham:2019b}.} given by
\begin{IEEEeqnarray}{rCl}
\mathcal{M}_\bullet(M_{\rm NC,\star})=&(&2.62\pm0.42)\log\left(\frac{M_{\rm NC,\star}}{10^{7.83}\,\mathrm{M_{\odot}}}\right)+\nonumber\\
&(&8.22\pm0.20),
\label{eqn:nc_rel}
\end{IEEEeqnarray}
and with an intrinsic scatter of 1.31\,dex. However, due to AGN contamination, we treat this as an upper-limit black hole mass estimate. Therefore, we predict the following black hole mass,
\begin{equation}
\mathcal{M}_{\rm NC,\star}\leq6.41\pm0.16\rightarrow \mathcal{M}_\bullet(M_{\rm NC,\star})\leq4.51\pm1.51,
\label{eqn:nc}
\end{equation}
with $P(\mathcal{M}_\bullet\leq5)\geq63\%$.

\subsection{The $M_\bullet$--$v_{\rm rot}$ relation}

From HyperLeda\footnote{\url{http://leda.univ-lyon1.fr/}} \citep{Paturel:2003}, we adopted their apparent maximum rotation velocity of the gas, $v_{\rm max,g}=84.33\pm1.80\,{\rm km\,s^{-1}}$ (homogenised value derived from 24 independent measurements), which is the observed maximum rotation velocity uncorrected for inclination effect. We then converted this to a maximum physical velocity rotation corrected for inclination ($v_{\rm rot}$) via
\begin{equation}
v_{\rm rot} \equiv \frac{v_{\rm max,g}}{\sin{i_{\rm disk}}} = 102.21\pm2.20\,{\rm km\,s^{-1}}.
\label{eqn:vmaxg}
\end{equation}
Application of equation~10 from \citet{Davis:2019b} gives
\begin{equation}
v_{\rm rot}=102.21\pm2.20\,{\rm km\,s^{-1}}\rightarrow \mathcal{M}_\bullet(v_{\rm rot})=3.90\pm0.59,
\label{eqn:vrot}
\end{equation}
with $P(\mathcal{M}_\bullet\leq5)=97\%$.

\subsection{The $M_\bullet$--$\sigma_0$ relation}\label{sec:msigma}

We obtained the central stellar velocity dispersion from \citet{Ho:2009} and utilised equation~2 from \citet{Sahu:2019b} to predict a black hole mass as follows,
\begin{equation}
\sigma_0=87.4\pm9.2\,{\rm km\,s^{-1}}\rightarrow \mathcal{M}_\bullet(\sigma_0)=6.08\pm0.67,
\label{eqn:sigma}
\end{equation}
with $P(\mathcal{M}_\bullet\leq5)=5\%$. This black hole mass estimate is the highest of all our estimates; it is our only discrete mass estimate of NGC~3319* with $\mathcal{M}_\bullet>5.2$.

\citet{Ho:2009} presented a catalogue of pre-existing velocity dispersions, observed sometime between 1982 and 1990 \citep{Ho:1995}. The measurements were weighted-mean dispersions from the blue- and red-side of the Double Spectrograph \citep{Oke:1982} mounted at the Cassegrain focus of the Hale 5.08\,m telescope at Palomar Observatory. However, \citet{Ho:2009} found that the blue-side spectral resolution is insufficient to reliably measure dispersions for most of the later-type galaxies in their sample, as was the case for NGC~3319. \citet{Ho:2009} only presented a red-side velocity dispersion for NGC~3319. Moreover, \citet{Ho:1995} noted that for their observations of NGC~3319 `the continuum shape of its spectrum may be uncertain because of imperfect correction for spatial focus variations'.

Although \citet{Jiang:2018} do present a spectrum of NGC~3319 (see their figure~5) from the \textit{Sloan Digital Sky Survey (SDSS)}, they do not report on the velocity dispersion. The \textit{SDSS} Data Release 12 \citep{Alam:2015} states\footnote{\url{https://www.sdss.org/dr12/algorithms/redshifts/}} that `best-fit velocity-dispersion values $\lesssim$100\,km\,s$^{-1}$ are below the resolution limit of the \textit{SDSS} spectrograph and are to be regarded with caution'. Nonetheless, we have attempted to measure the velocity dispersion from the \textit{SDSS} spectrum (Fig.~\ref{fig:spec}) and found $\sigma_0 = 99\pm9\,{\rm km\,s^{-1}}$ (given the aforementioned resolution limit, this is likely an upper limit), albeit with a discrepant estimate of its recessional velocity. We found $cz=860\pm6\,{\rm km\,s^{-1}}$, which is markedly different from the \textit{SDSS} value ($cz=713\pm5\,{\rm km\,s^{-1}}$), or even the mean heliocentric radial velocity from HyperLeda ($cz=738\pm7\,{\rm km\,s^{-1}}$). Although $\sigma_0\lesssim100$\,km\,s$^{-1}$, and is thusly suspicious, our measurement of $\sigma_0 = 99\pm9\,{\rm km\,s^{-1}}$ is consistent with the value from \citep{Ho:1995}. Better spectral resolution should provide greater clarity as to the velocity dispersion of this galaxy, which might also be influenced by the nuclear star cluster.

\begin{figure}
\includegraphics[clip=true, trim= 1mm 1.25mm 1.25mm 1.5mm, width=\columnwidth]{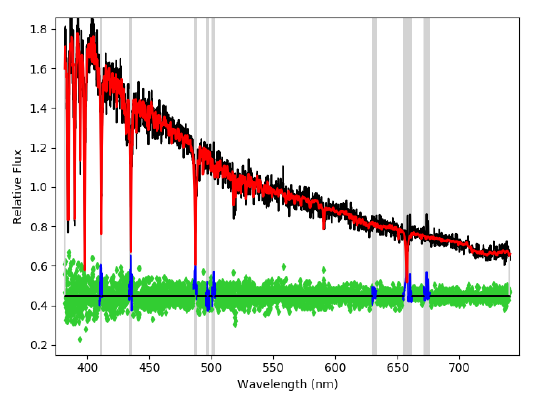}
\caption{Fit to the \textit{SDSS} spectrum of NGC~3319 by \textsc{ppxf} \citep{Cappellari:2017}. The relative flux of the observed spectrum (\hdashrule[0.35ex]{6mm}{0.4mm}{}) is over-plotted by the \textsc{ppxf} fit (\textcolor{red}{\hdashrule[0.35ex]{6mm}{0.4mm}{}}) to the spectrum. The residuals to the fit (\textcolor{LimeGreen}{$\blacklozenge$}) are at the bottom (normalised about the arbitrary horizontal black line) along with residuals from the masked emission features (\textcolor{blue}{\hdashrule[0.35ex]{6mm}{0.4mm}{}}), while grey vertical bands delineate the masked regions not included in the $\chi^2$ minimisation of the fit. The fit is consistent with $\sigma_0 = 99\pm9\,{\rm km\,s^{-1}}$ and $cz=860\pm6\,{\rm km\,s^{-1}}$.}
\label{fig:spec}
\end{figure}

\subsection{The $M_\bullet$--$L_{\rm 2-10\,keV}$ relation}\label{sec:Lx}

\citet{Mayers:2018} studied a sample of 30 AGN ($z\leq0.23$, $L_{\rm 2-10\,keV}\geq10^{40.8}\,{\rm erg\,s^{-1}}$, and $\mathcal{M}_\bullet\geq5.45$), with black hole masses estimated via \citet{Bentz:2015} and reported a trend between black hole mass and X-ray luminosity. From 2-10\,keV \textit{CXO} observations of the nuclear point source in NGC~3319, \citet{Jiang:2018} calculated a luminosity of $L_{\rm 2-10\,keV}=10^{39.0\pm0.1}\,{\rm erg\,s^{-1}}$. We applied the $M_\bullet$--$L_{\rm 2-10\,keV}$ relation of \citet[][extracted from their figure~11]{Mayers:2018},
\begin{IEEEeqnarray}{rCl}
\mathcal{M}_\bullet(L_{\rm 2-10\,keV}) = &(&0.58\pm0.05)\log\left(\frac{L_{\rm 2-10\,keV}}{2\times10^{43}\,{\rm erg\,s^{-1}}}\right)+\nonumber\\
&(&7.46\pm0.34),
\label{eqn:Mayers}
\end{IEEEeqnarray}
with a scatter of 0.89\,dex, such that
\begin{IEEEeqnarray}{rCl}
L_{\rm 2-10\,keV}&=&10^{39.0\pm0.1}\,{\rm erg\,s^{-1}}\rightarrow\nonumber\\
\mathcal{M}_\bullet(L_{\rm 2-10\,keV})&=&4.97\pm0.98,
\label{eqn:Lx}
\end{IEEEeqnarray}
with $P(\mathcal{M}_\bullet\leq5)=51\%$. However, given that the Eddington ratio will vary over time, as the AGN duty cycle turns the AGN on and off, this is unlikely to be a stable mass estimate.\footnote{\citet{Woo:2002} found that the Eddington ratio for a given black hole can vary, spanning a range of up to three orders of magnitude. In order to be a stable relation, the $M_\bullet$--$L_{\rm 2-10\,keV}$ relation would require the time-varying distribution of Eddington ratios for a given black hole to resemble a normal distribution; several studies have found supportive evidence for a peaked distribution \citep{Kollmeier:2006,Steinhardt:2010,Lusso:2012}.} The fundamental plane of black hole activity (Section~\ref{sec:FP}) can offer additional insight, with its counterbalance from the waxing/waning radio emission.\footnote{Although, unmatched (in the radio) X-ray variability \citep[typically not more than a factor of $\approx$3;][]{Timlin:2020} can possibly contribute to the scatter in the relation.}

In what follows (Sections \ref{sec:nxs}--\ref{sec:FP}) are three black hole mass estimates from \citet{Jiang:2018}, which are explicitly described here.

\subsection{The $M_\bullet$--$\sigma_{\rm NXS}^2$ relation}\label{sec:nxs}

From the light curves obtained by the \textit{CXO} observations, \citet{Jiang:2018} estimated the X-ray variability, represented as the (10\,ks) normalised excess variance ($\sigma_{\rm NXS}^2$). They found $\sigma_{\rm NXS}^2=0.093\pm0.088$. By applying the $M_\bullet$--$\sigma_{\rm NXS}^2$ relation from \citet[][figure~4]{Pan:2015}, \citet{Jiang:2018} obtained
\begin{equation}
\sigma_{\rm NXS}^2=0.093\pm0.088\rightarrow \mathcal{M}_\bullet(\sigma_{\rm NXS}^2)=5.18\pm1.92,
\label{eqn:NXS}
\end{equation}
with $P(\mathcal{M}_\bullet\leq5)=46\%$. Clearly, with an upper 1$\sigma$ estimate for the black hole mass of $\sim$10$^7$\,M$_{\odot}$, on its own this is not evidence for an IMBH.

\subsection{Eddington ratio}\label{sec:Edd}

Based upon the median radio-quiet quasar spectral energy distribution (SED) of \citet{Elvis:1994}, \citet{Jiang:2018} determined a bolometric luminosity of $L_{\rm bol}=(3.6\pm1.1)\times10^{40}\,{\rm erg\,s^{-1}}$ for NGC~3319*, by scaling the SED to the \textit{CXO} luminosity ($L_{\rm 2-10\,keV}=10^{39.0\pm0.1}\,{\rm erg\,s^{-1}}$) and integrating the entire SED. Using \textit{XMM-Newton} and \textit{CXO} observations of NGC~3319*, \citet{Jiang:2018} also determined a hard X-ray photon index of $\Gamma=2.02\pm0.27$. Following \citet{Jiang:2018}, we converted this into an Eddington ratio, $\log(L_{\rm bol}/L_{\rm Edd})=-0.56\pm0.99$, with an Eddington luminosity, $L_{\rm Edd}\equiv1.26\times10^{38}\,M_\bullet(\rm M_{\odot}^{-1}\,{erg\,s}^{-1})$, via equation~2 from \citet{Shemmer:2008}. Therefore, $L_{\rm Edd}=10^{41.12\pm1.00}\,{\rm erg\,s^{-1}}$.

From this point in the calculation, \citet{Jiang:2018} arbitrarily selected $L_{\rm bol}/L_{\rm Edd}=0.1^{+0.9}_{-0.099}$, implying $M_\bullet=3^{+297}_{-2.7}\times10^3\,\mathrm{M_\odot}$. Thus, \citet{Jiang:2018} broadened the mass estimate to a range from $M_\bullet=3\times10^2$ to $3\times10^5\,\mathrm{M_{\odot}}$ for arbitrary Eddington ratios ranging from 1 to $10^{-3}$, a range of 3\,dex. For our purposes, we will remain with the calculated $\log(L_{\rm bol}/L_{\rm Edd})=-0.56$ with $L_{\rm Edd}=10^{41.12}\,{\rm erg\,s^{-1}}$, but will follow \citet{Jiang:2018}'s conservative 3\,dex range of uncertainty by broadening our estimate to
\begin{equation}
L_{\rm Edd}=10^{41.12\pm1.50}\,{\rm erg\,s^{-1}}\rightarrow \mathcal{M}_\bullet(L_{\rm Edd})=3.02\pm1.50,
\label{eqn:Edd}
\end{equation}
with $P(\mathcal{M}_\bullet\leq5)=91\%$.

\subsection{Fundamental plane of black hole activity}\label{sec:FP}

\citet{Baldi:2018} obtained high-resolution ($\leq$$0\farcs2$) 1.5\,GHz-radio images of the core in NGC~3319 but failed to detect a source; therefore, establishing an upper limit to the luminosity, $L_{1.5\,{\rm GHz}}\leq10^{35.03}\,{\rm erg\,s}^{-1}$.\footnote{\citet{Baldi:2018} presented $L_{1.5\,{\rm GHz}}\leq10^{34.84}\,{\rm erg\,s}^{-1}$ for NGC~3319, based on their adopted distance of 11.5\,Mpc.} This radio luminosity can be applied to the fundamental plane of black hole activity \citep{Merloni:2003,Falcke:2004,Gultekin:2009,Plotkin:2012,Dong:2015,Liu:2016,Nisbet:2016}, which demonstrates an empirical correlation between the continuum X-ray, radio emission, and mass of an accreting black hole. This fundamental plane applies to supermassive, as well as stellar-mass black holes; therefore, it should also be suitable for the intervening population of IMBHs \citep[e.g.][]{Gultekin:2014}. Using the fundamental plane of black hole activity, \citet{Jiang:2018} reported a black hole mass estimate of $\leq$$10^5\,\mathrm{M_{\odot}}$. However, it is typically the 5\,GHz, not the 1.5\,GHz luminosity as we have, that is employed in the fundamental plane relation. Therefore, we follow the radiative flux density, $S_\nu\propto\nu^{\alpha_R}$, conversion of \citet{Qian:2018} by adopting $\alpha_R=-0.5\pm0.1$ as the typical radio spectral index of bright (high Eddington ratio) AGN. Doing so, this predicts an associated 5\,GHz luminosity of $L_{5\,{\rm GHz}}\leq10^{34.77\pm0.05}\,{\rm erg\,s}^{-1}$. Using this value along with $L_{\rm 2-10\,keV}=10^{39.0\pm0.1}\,{\rm erg\,s^{-1}}$ (Section~\ref{sec:Lx}), we applied the relation of \citet[][equation~8]{Gultekin:2019} to predict the following upper limit to the black hole mass,
\begin{IEEEeqnarray}{rCl}
L_{\rm FP}&\equiv&\left(L_{\rm 2-10\,keV},L_{5\,{\rm GHz}}\right)\nonumber \\ 
&=& (10^{39.0\pm0.1},{\rm \leq}10^{34.77\pm0.05})\,{\rm erg\,s}^{-1}\rightarrow\nonumber \\ 
\mathcal{M}_\bullet(L_{\rm FP})&\leq&5.62\pm1.05,
\label{eqn:FP}
\end{IEEEeqnarray}
with $P(\mathcal{M}_\bullet\leq5)\geq28\%$.\footnote{Given the connection between the black hole mass estimates from the Eddington ratio (equation~\ref{eqn:Edd}) and the fundamental plane (equation~\ref{eqn:FP}) we also check that the former ($L_{\rm 2-10\,keV}=10^{39.0\pm0.1}\,{\rm erg\,s^{-1}}$ and $\mathcal{M}_\bullet(L_{\rm Edd})=3.02\pm1.50$) is consistent with no radio detection ($L_{5\,{\rm GHz}}\leq10^{34.77\pm0.05}\,{\rm erg\,s}^{-1}$). Using equation~19 from \citet{Gultekin:2019}, with the radio luminosity as the dependent variable in their regression, we find $L_{5\,{\rm GHz}}=10^{33.76\pm1.41}\,{\rm erg\,s}^{-1}$. Thus, the inverse prediction is consistent with no radio detection.}

However, two issues make this particular prediction problematic. The first is that the radio and X-ray data were not obtained simultaneously, and the timescale for variations in flux will be short for IMBHs given that it scales with the size of the ‘event horizon’ and thus with the black hole mass. The second issue is that the ‘fundamental plane of black hole activity’ is applicable to black holes with low accretion rates \citep[][their final paragraph of section~6]{Merloni:2003}, and NGC 3319* is considered to have a high accretion rate \citep[][see their section~3.2]{Jiang:2018}.  Therefore, we do not include this estimate in our derivation of the black hole mass.

\subsection{The $M_\bullet$--$\mathcal{C}_\text{FUV,tot}$ relation}\label{sec:colour}

\citet{Dullo:2020} present a relationship between the black hole mass and its host galaxy's UV$-3.6\,\micron$ colour\footnote{See also the dependence of black hole mass on the colour of its host galaxy presented by \citet{Zasov:2013}.} from their study of 67 galaxies with directly-measured black hole masses. From table~D1 in \citet{Dullo:2020}, the predicted black hole mass for NGC~3319* is $\mathcal{M}_\bullet=5.36\pm0.85$, based on its FUV$-3.6\,\micron$ colour ($\mathcal{C}_\text{FUV,tot}$).\footnote{\citet{Dullo:2020} also supply an $M_\bullet$--$\mathcal{C}_\text{NUV,tot}$ relation, but given the similarity with the $M_\bullet$--$\mathcal{C}_\text{FUV,tot}$ relation, we prefer to use the FUV relation due to its smaller uncertainty on the slope and intercept.} However, we can further refine this prediction by accounting for the internal dust extinction in NGC~3319. Given that NGC~3319 is bulgeless, we treat it as being all disk. Using our adopted inclination, $i_{\rm disk}=55\fdg6\pm0\fdg2$, and applying equations 2 and 4 from \citet[][see also \citealt{Driver:2008}]{Dullo:2020}, we find that these corrections make NGC~3319 $0.57\pm0.16$\,mag brighter in the ultraviolet and $0.05\pm0.02$\,mag brighter at $3.6\,\micron$.\footnote{We note the caveat that the relations of \citet{Dullo:2020} are based on bulge plus disk magnitudes, not total galaxy magnitudes. In contrast, the colours from \citet{Bouquin:2018}, which were used to predict black hole masses in table~D1 of \citet{Dullo:2020}, are derived from asymptotic magnitudes that may include additional fluxes from bars, rings, and nuclear components. For NGC~3319, we assume the bar and disk have the same colour and require the same dust correction because bars are just the inner parts of disks that have changed their orbits.} Thus, the change in colour will be $0.52\pm0.14$\,mag bluer, which updates the FUV$-3.6\,\micron$ colour from \citet{Bouquin:2018} to an internal-dust-corrected $\mathcal{C}_\text{FUV,tot}=1.16\pm0.14$\,mag. Using the BCES bisector \citep{BCES} $M_\bullet$--$\mathcal{C}_\text{FUV,tot}$ relation for late-type galaxies with a slope of $1.03\pm0.13$ from table~2 in \citet{Dullo:2020}, we obtain a black hole mass which is $1.03\pm0.13\times0.52\pm0.14=0.53\pm0.16$\,dex smaller than reported in table~D1. This revision reduces the tabulated estimate of $\mathcal{M}_\bullet$ from $5.36\pm0.85$ to $4.83\pm0.87$. Thus,
\begin{IEEEeqnarray}{rCl}
\mathcal{C}_\text{FUV,tot}&=&1.16\pm0.14\,{\rm mag}\rightarrow\nonumber\\
\mathcal{M}_\bullet(\mathcal{C}_\text{FUV,tot})&=&4.83\pm0.87,
\label{eqn:colour}
\end{IEEEeqnarray}
with $P(\mathcal{M}_\bullet\leq5)=58\%$.

\subsection{Probability density function}\label{sec:PDF}

With such a multitude of mass estimates and a hesitancy to place confidence in one measurement alone, we combined the aforementioned mass estimates (except for that from equation~\ref{eqn:Lx}) to yield a single black hole mass estimate for NGC~3319*. We did so by analysing the probability density function (PDF) of the distribution of mass estimates (see Fig.~\ref{fig:PDF}). For our seven selected black hole mass estimates (equations \ref{eqn:phi_wm}, \ref{eqn:Mgal}, \ref{eqn:vrot}, \ref{eqn:sigma}, \ref{eqn:NXS}, \ref{eqn:Edd}, and \ref{eqn:colour}), we let a normal distribution represent each estimate with their respective means ($\mathcal{\bar{M}}_\bullet$) and standard deviations ($\delta\mathcal{\bar{M}}_\bullet$). We then added the seven Gaussians together to produce a combined summation. To ensure the area of the summation is equal to one, we divided each of the seven Gaussian addends by seven so that the area under each Gaussian equalled $1/7$.

\begin{figure}
\includegraphics[clip=true, trim= 0mm 0mm 0mm 0mm, width=\columnwidth]{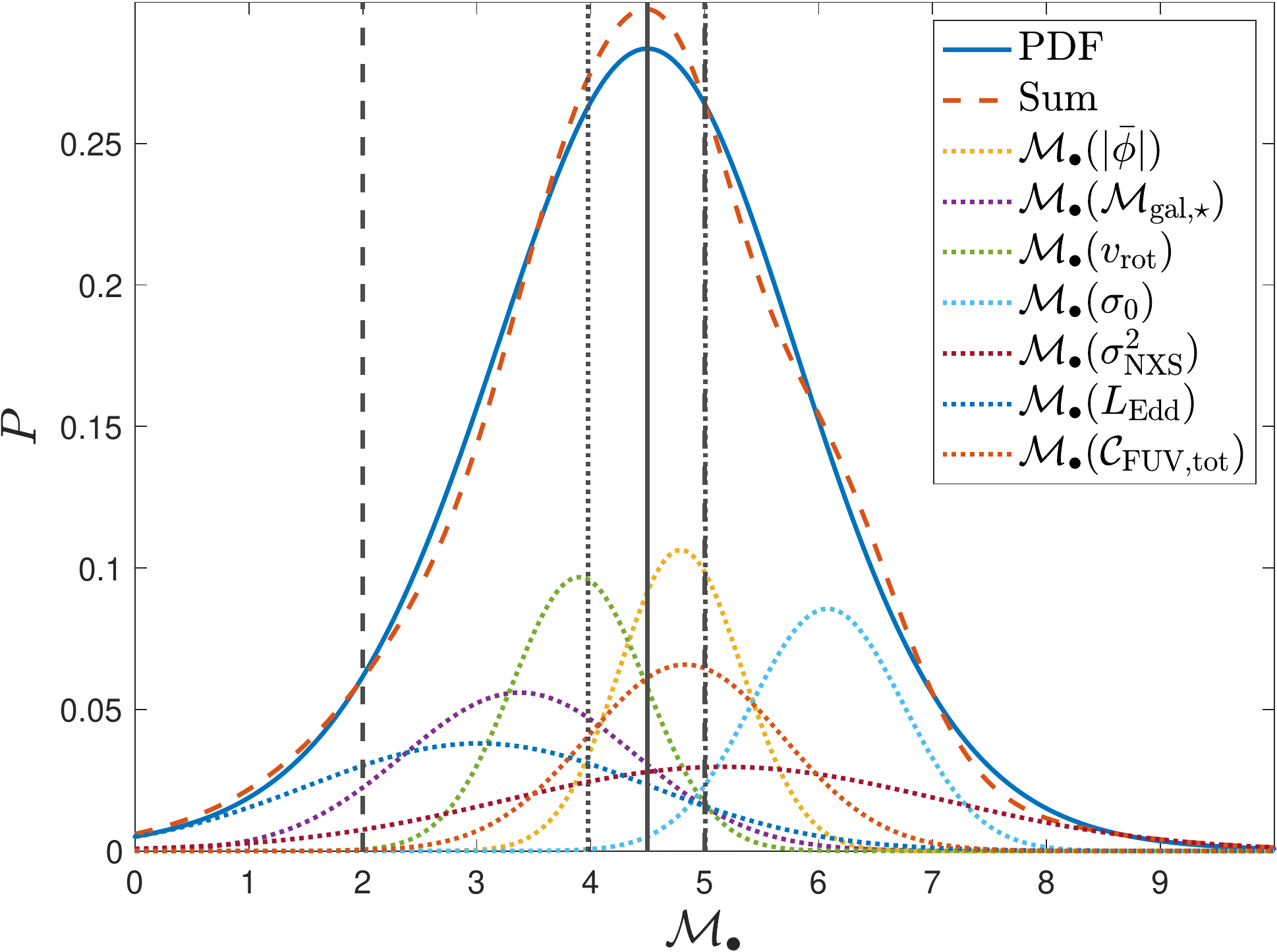}
\caption{
Determination of the PDF of the black hole mass estimates for NGC~3319*. The PDF (\textcolor{MATLAB_blue}{{\hdashrule[0.35ex]{8mm}{0.4mm}{}}}) is the best-fit skew-kurtotic-normal distribution to the Sum (\textcolor{MATLAB_red}{{\hdashrule[0.35ex]{8mm}{1pt}{1mm}}}) of each of the seven selected black hole mass estimates' normal distributions;\\
\textcolor{MATLAB_red}{{\hdashrule[0.35ex]{8mm}{1pt}{1mm}}} = \textcolor{MATLAB_orange}{{\hdashrule[0.35ex]{8mm}{1pt}{1pt}}} + \textcolor{MATLAB_purple}{{\hdashrule[0.35ex]{8mm}{1pt}{1pt}}} + \textcolor{MATLAB_green}{{\hdashrule[0.35ex]{8mm}{1pt}{1pt}}} + \textcolor{MATLAB_cyan}{{\hdashrule[0.35ex]{8mm}{1pt}{1pt}}} + \textcolor{MATLAB_maroon}{{\hdashrule[0.35ex]{8mm}{1pt}{1pt}}} + \textcolor{MATLAB_blue}{{\hdashrule[0.35ex]{8mm}{1pt}{1pt}}} + \textcolor{MATLAB_red}{{\hdashrule[0.35ex]{8mm}{1pt}{1pt}}}.
The solid vertical line ({\hdashrule[0.35ex]{8mm}{0.4mm}{}}) indicates the position of\\
$\mathcal{\widehat{M}}_\bullet$. The dotted vertical lines ({\hdashrule[0.35ex]{8mm}{1pt}{1pt}}) demarcate $\mathcal{\widehat{M}}_\bullet-\delta^-\mathcal{\widehat{M}}_\bullet$ (left) and $\mathcal{\widehat{M}}_\bullet+\delta^+\mathcal{\widehat{M}}_\bullet$ (right, overlapping with the dashed line). The dashed vertical lines ({\hdashrule[0.35ex]{8mm}{1pt}{1mm}}) demarcate the upper- and lower-bound mass definitions of an IMBH.
}
\label{fig:PDF}
\end{figure}

We fit a skew-kurtotic-normal distribution to the summation and measured the peak (mode) black hole mass of the PDF as 
\begin{equation}
\mathcal{\widehat{M}}_\bullet \equiv \mathcal{M}_\bullet(\max{P}) = 4.50_{-0.52}^{+0.51},
\label{eqn:Mhat}
\end{equation}
where $\mathcal{M}_\bullet(\max{P})$ is the black hole mass when the probability ($P$) reaches its maximum ($\max{P}=0.272$). We quantify its standard error as
\begin{IEEEeqnarray}{rCl}
\delta^+\mathcal{\widehat{M}}_\bullet &\equiv& \frac{{\rm RWHM}}{\sqrt{2N\ln{2}}} = 0.51\ \ \ \mathrm{and}\nonumber\\
\delta^-\mathcal{\widehat{M}}_\bullet &\equiv& \frac{{\rm LWHM}}{\sqrt{2N\ln{2}}} = 0.52,
\label{eqn:sem}
\end{IEEEeqnarray}
with right width at half max ${\rm RWHM}=1.59$\,dex, left width at half max ${\rm LWHM}=1.63$\,dex, the number of predictors $N=7$, and $P(\mathcal{\widehat{M}}_\bullet\leq5)=84\%$. For a complete comparison of all the mass estimates, see Table~\ref{tab:masses} and Fig.~\ref{fig:forest}.

\begin{table*}
\centering
\caption{NGC~3319* mass predictions.
\textbf{Columns:}
{\bf (1)} Black hole mass scaling relation predictor.
{\bf (2)} Logarithmic black hole (solar) mass.
{\bf (3)} Probability that NGC~3319* is $\leq$$10^5\,\mathrm{M_{\odot}}$, via equation~(\ref{eqn:prob}).
}
\label{tab:masses}
\begin{tabular}{lrr}
\hline
Predictor & $\mathcal{M}_\bullet\pm\delta\mathcal{M}_\bullet$ & $P(\mathcal{M}_\bullet\leq5)$ \\
 & (dex) & (\%) \\
(1) & (2) & (3) \\
\hline
$|\bar{\phi}|=28\fdg0\pm2\fdg3$ & $4.79\pm0.54$ & 65 \\
$\mathcal{M}_{{\rm gal},\star}=9.53\pm0.10$ & $3.38\pm1.02$ & 94 \\
$\mathcal{M}_{\rm NC,\star}\leq6.41\pm0.16$ & $\leq$$4.51\pm1.51^\dagger$ & $\geq$63 \\
$v_{\rm rot}=102.21\pm2.20\,{\rm km\,s^{-1}}$ & $3.90\pm0.59$ & 97 \\
$\sigma_0=87.4\pm9.2\,{\rm km\,s^{-1}}$ & $6.08\pm0.67$ & 5 \\
$L_{\rm 2-10\,keV}=10^{39.0\pm0.1}\,{\rm erg\,s^{-1}}$ & $4.97\pm0.98^\dagger$ & 51 \\
$\sigma_{\rm NXS}^2=0.093\pm0.088$ & $5.18\pm1.92$ & 46 \\
$L_{\rm Edd}=10^{41.12\pm1.50}\,{\rm erg\,s^{-1}}$ & $3.02\pm1.50$ & 91 \\
${L_{\rm FP}}\equiv\left(L_{\rm 2-10\,keV},L_{5\,{\rm GHz}}\right)=(10^{39.0\pm0.1},$$\leq$$10^{34.77\pm0.05})\,{\rm erg\,s}^{-1}$ & $\leq$$5.62\pm1.05^\dagger$ & $\geq$28 \\
$\mathcal{C}_\text{FUV,tot}=1.16\pm0.14$\,mag & $4.83\pm0.87$ & 58 \\
\textbf{PDF} & \bm{$4.50_{-0.52}^{+0.51}$} & \textbf{84} \\
\hline
\multicolumn{3}{l}{$^\dagger$ Excluded from the black hole mass PDF.} \\
\end{tabular}
\end{table*}

\begin{figure}
\includegraphics[clip=true, trim= 0mm 0mm 0mm 0mm, width=\columnwidth]{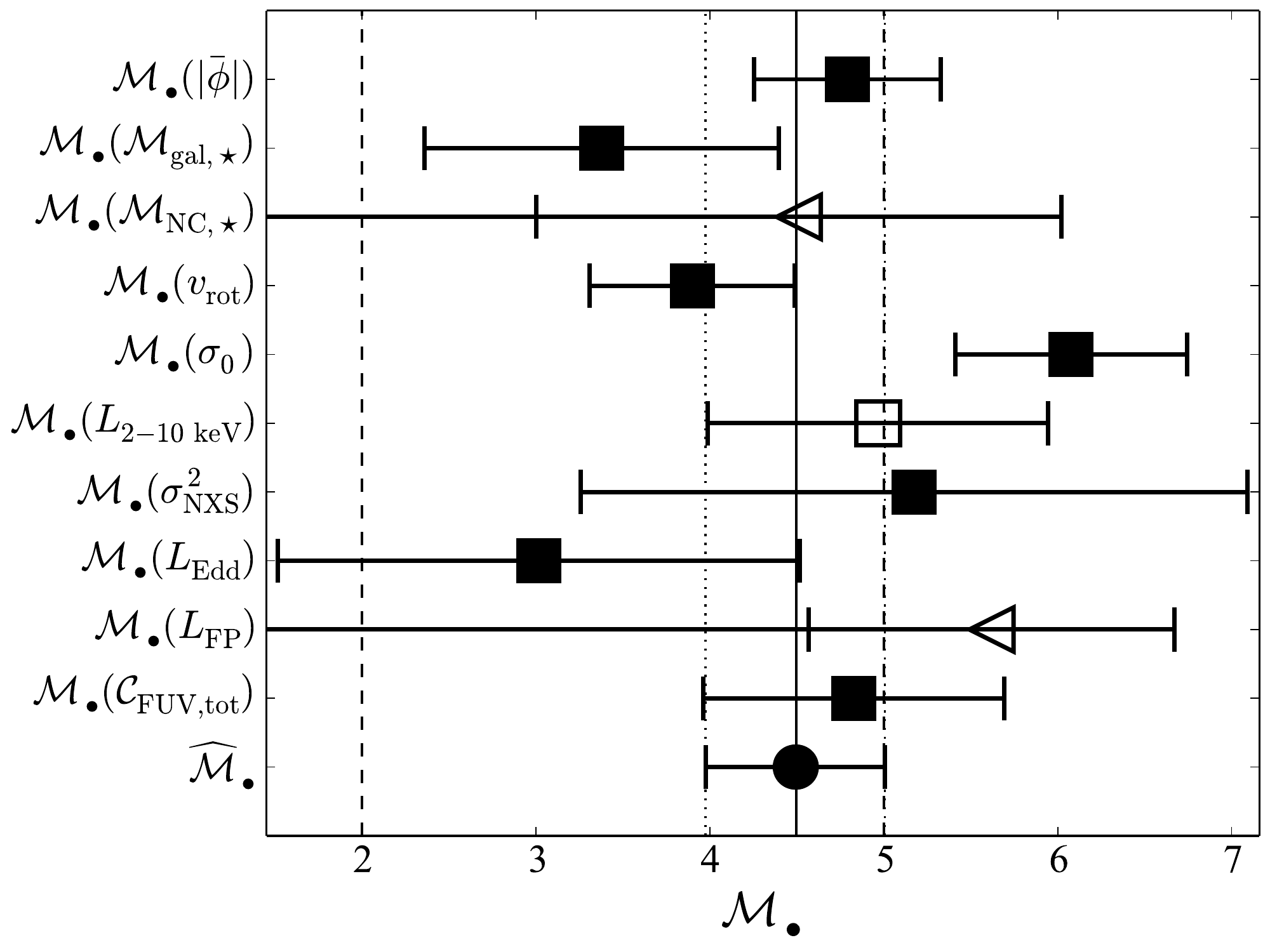}
\caption{
Forest plot of the ten different black hole mass estimates of NGC~3319*. Seven discrete mass estimates ($\blacksquare$) are used to generate the $\mathcal{\widehat{M}}_\bullet$ estimate ($\bullet$) at the bottom of the figure. The black hole mass estimate from the X-ray luminosity, $\mathcal{M}_\bullet(L_{2-10\,{\rm keV}})$, is plotted ($\square$), but not included in the calculation of $\mathcal{\widehat{M}}_\bullet$. Nor are the upper-limit black hole mass estimates included in the calculation of $\mathcal{\widehat{M}}_\bullet$. The black hole mass estimates from the nuclear star cluster mass, $\mathcal{M}_\bullet(\mathcal{M}_{\rm NC,\star})$, and the fundamental plane of black hole activity, $\mathcal{M}_\bullet(L_{\rm FP})$, are upper-limit black hole mass estimates depicted by left-pointing open triangles ($\triangleleft$). The vertical lines are equivalent to those found in Fig.~\ref{fig:PDF}.
}
\label{fig:forest}
\end{figure}

\section{Discussion}\label{sec:disc}

We have presented multiple mass estimates for NGC~3319*, eight of which are discrete estimates, and two are upper limits (Sections \ref{sec:NC} and \ref{sec:FP}). The non-detection of a nuclear source in the radio observations places an upper limit that is indeed higher than most of our other mass estimates. This missing radio detection begs for future deep, high spatial resolution radio (along with simultaneous X-ray) observations to provide an improved mass estimate for NGC~3319* via the fundamental plane of black hole activity. Nonetheless, the upper-limit mass estimate from the fundamental plane of black hole activity (equation~\ref{eqn:FP}) is in agreement with our other mass estimate derived from X-ray measurements (equation~\ref{eqn:Lx}).

Amongst our numerous mass estimates, it is perhaps the most well-known black hole mass scaling relation ($M_\bullet$--$\sigma_0$) that produces the highest mass estimate. Indeed, equation~(\ref{eqn:sigma}) provides the only mass estimate that is not consistent with $\mathcal{M}_\bullet\leq5$. It would be of interest to obtain a suitably high-spectral resolution measurement of $\sigma_0$ for NGC~3319 to confirm or revise the solitary measurement that is (now) at least 31 years old. Although, it is not unprecedented to find a black hole that is anomalously under-massive with respect to the $M_\bullet$--$\sigma_0$ relation \citep{Zaw:2020}.

We have used the latest refinement of the $M_\bullet$--$\sigma_0$ relation by \citet{Sahu:2019b} to estimate the black hole mass. Building on \citet{Davis:2017}, \citet{Sahu:2019b} have determined that $M_\bullet\propto\sigma_0^{5.82\pm0.75}$ from an analysis of 46 spiral galaxies with central velocity dispersion measurements and directly-measured black hole masses. However, none of these galaxies have black hole masses below $\sigma_0 \approx 100$\,km\,s$^{-1}$ ($M_\bullet = 10^6\,\mathrm{M_\odot}$). \citet[][their figure~2]{Sahu:2019b} have revealed a tendency for galaxies with central velocity dispersions less than $\sim$100\,km\,s$^{-1}$ to reside above the $M_\bullet$--$\sigma_0$ relation defined by the galaxies with higher velocity dispersions and directly-measured black hole masses (i.e. spatially resolved kinematics, not reverberation mapping nor single-epoch spectra coupled with a constant virial $f$-factor). Therefore, should the velocity dispersion of NGC~3319 be lower than $\sigma_0 \approx 100$\,km\,s$^{-1}$, a shallower $M_\bullet$--$\sigma_0$ relation than used here will be required. 

\citet{Baldassare:2020} demonstrated that extrapolations of the shallow $M_\bullet$--$\sigma_0$ relation for `classical bulges' from \citet{Kormendy:2013} appears (perhaps superficially) valid down to black hole masses of $10^5\,\mathrm{M_{\odot}}$, with black hole mass estimates derived from single-epoch spectroscopic (virial; with assumption of an $f$-factor to account for the unknown broadline region geometry) masses. If we exclude the $M_\bullet$--$\sigma_0$ mass estimate altogether, our $\mathcal{\widehat{M}}_\bullet$ black hole mass estimate for NGC~3319* (equation~\ref{eqn:Mhat}) becomes $\mathcal{\widehat{M}}_\bullet = 4.14_{-0.49}^{+0.50}$, with $P(\mathcal{\widehat{M}}_\bullet\leq5)=96\%$, based on the remaining six discrete measures used in Fig.~\ref{fig:PDF}. Additionally, if we treat the nuclear star cluster upper-limit mass estimate as a discrete estimate, we arrive at $\mathcal{\widehat{M}}_\bullet = 4.19_{-0.47}^{+0.48}$, also with $P(\mathcal{\widehat{M}}_\bullet\leq5)=96\%$. This is based again on seven measures, except now excluding the $M_\bullet$--$\sigma_0$ and $M_\bullet$--$L_X$ relation estimates, as well as the fundamental plane estimate.

\subsection{Similarity to the IMBH in LEDA~87300}

The IMBH candidate in LEDA~87300 (RGG 118) has been proclaimed the `smallest' reported in a galaxy nucleus\footnote{The first strong IMBH candidate is an off-centre ultra-luminous X-ray point source that is too bright to be an accreting stellar-mass black hole \citep{Farrell:2009}.} \citep{Baldassare:2015,Baldassare:2017,Graham:2016b}. We adopt the same redshift ($z=0.02647\pm0.00026$) as \citet{Graham:2016b}, but instead invoke the latest cosmographic parameters ($H_0 = 67.66\pm0.42$\,km\,s$^{-1}$\,Mpc$^{-1}$, $\Omega_\Lambda=0.6889\pm0.0056$, and $\Omega_{\rm m}=0.3111\pm0.0056$) from \citet[][equation~28]{Planck:2018} to calculate a Hubble flow (comoving radial) distance of $116.6\pm1.3$\,Mpc \citep{Wright:2006}. This adjustment yields a mass of $\mathcal{M}_\bullet=4.48^{+0.52}_{-0.69}$ for the IMBH (LEDA~87300*) in LEDA~87300, as determined by \citet{Graham:2016b}, with $P(\mathcal{M}_\bullet\leq5)=84\%$. This was based on a virial $f$-factor of 2.8 \citep{Graham:2011} and the assumption that the $M_\bullet$--$\sigma_0$ relation for AGN and quiescent galaxies can be extrapolated below $10^6$\,M$_\odot$. Thus, the masses of NGC~3319* and LEDA~87300* are nearly identical, $3.14_{-2.20}^{+7.02}\times10^4\,\mathrm{M_\odot}$ and $3.00_{-2.38}^{+6.93}\times10^4\,\mathrm{M_\odot}$, respectively. However, given the overlapping error bars associated with both black holes, the best we can conclude at this time is that their masses may be similar.

\subsection{Environment and secular evolution}

NGC~3319 is a relatively isolated galaxy in a group of four galaxies: NGC~3104, 3184, 3198, and 3319 \citep{Tully:1988}. Its nearest neighbour at present is most likely NGC~3198. NGC~3198 is at a distance ($d$) from us of $14.5\pm1.3$\,Mpc \citep[Cepheid variable star distance from][]{Kelson:1999}, J2000 right ascension ($\alpha$) of $10^{\rm h}19^{\rm m}55^{\rm s}$, and J2000 declination ($\delta$) of $+45\degr33\arcmin09\arcsec$, while NGC~3319 is at $d=14.3\pm1.1$\,Mpc, $\alpha=10^{\rm h}39^{\rm m}09\fs8$, and $\delta=+41\degr41\arcmin15\farcs9$. Based on the heliocentric spherical coordinates of each galaxy, the physical distance between galaxies is
\begin{IEEEeqnarray}{rCl}
\left \| \vec{d_1}-\vec{d_2} \right \| &\equiv& \sqrt{{d_1}^2+{d_2}^2-2d_1d_2\cos(\alpha_1-\alpha_2)} \\
&& -\sqrt{2d_1d_2\sin\alpha_1\sin\alpha_2[\cos(\delta_1-\delta_2)-1]}.\nonumber
\label{eqn:sep}
\end{IEEEeqnarray}
The physical separation between NGC~3198 and NGC~3319 is thus $1.3\pm0.2$\,Mpc.

With this level of isolation, NGC~3319 will likely experience many gigayears of relative tranquillity, without any significant galaxy mergers. If so, NGC~3319* should continue to coevolve along with its host galaxy via secular accretion and feedback. There is no telling evidence that NGC~3319 has experienced a recent major merger. However, we do note that \citet{Moore:1998} detected a small system ($4.2\times10^7\,\mathrm{M_{\odot}}$), just $11\arcmin$ ($46\pm4$\,kpc) south of NGC~3319. \citet{Moore:1998} postulate that tidal interactions between this object and NGC~3319 likely explain the distorted spiral structure, H\,\textsc{i} tail, and velocity perturbations in the southern half of the galaxy.

\subsection{Direct measurements of NGC~3319*}

Stellar remnant black holes are thought to exist between the Tolman-Oppenheimer-Volkoff limit of $\approx$$2.17\,\mathrm{M_{\odot}}$ for cold, non-rotating neutron stars \citep{Tolman:1939,Oppenheimer:1939,Margalit:2017,Shibata:2017,Ruiz:2018,Rezzolla:2018} and $\lesssim$60--$80\,\mathrm{M_{\odot}}$ from the collapse of massive stars estimated from evolutionary models \citep{Belczynski:2010,Woosley:2017,Spera:2017}. Recent observations have found the least massive known black hole \citep[$\approx$$3.3\,\mathrm{M_{\odot}}$;][]{Thompson:2019}.\footnote{See also the recent $3.04\pm0.06\,\mathrm{M}_\odot$ black hole candidate \citep{Jayasinghe:2021}.} Over the past couple of years, black holes have been discovered that begin to surpass the low-mass definition of IMBHs: $84.4^{+15.8}_{-11.1}\,\mathrm{M_{\odot}}$ \citep{Abbott:2019} and $98^{+17}_{-11}$\,M$_\odot$ \citep{Zackay:2019}. The gravitational-wave signal GW190521 \citep{LIGO:2020} is consistent with the BH-collisional-creation of a $142^{+28}_{-16}$\,M$_\odot$ IMBH. Its properties and astrophysical implications \citep{LIGO:2020b} are further remarkable given the high confidence that at least one of its progenitors lay in the mass gap predicted by pair-instability supernova theory \citep{Woosley:2017}.\footnote{Alternatively, \citet{Roupas:2019} propose that black holes between 50 and 135\,$\mathrm{M_{\odot}}$ can form via rapid gas accretion in primordial dense clusters.}

The dwarf elliptical galaxy NGC~205 (M110), which is a satellite of the Andromeda Galaxy (M31), is presently the least massive nuclear black hole measured via direct methods. \citet{Nguyen:2019} estimated a black hole mass of $\mathcal{M}_\bullet=3.83_{-0.60}^{+0.43}$ via stellar dynamical modelling. Furthermore, this galaxy seemingly confirms the extrapolation of scaling relations into the IMBH regime. Explicitly, its black hole mass is consistent with the prediction, $\mathcal{M}_\bullet(\sigma_0)=3.86\pm0.55$, of the $M_\bullet$--$\sigma_0$ relation \citep[][equation~1]{Sahu:2019b} with $\sigma_0=33.1\pm4.8\,{\rm km\,s^{-1}}$ from HyperLeda.

In order to dynamically estimate the mass of NGC~3319*, it is necessary to resolve motions within its sphere of influence (SOI). According to \citet{Peebles:1972}, the gravitational SOI of a black hole residing at the centre of a galaxy has a radius,
\begin{equation}
r_h \equiv \frac{GM_\bullet}{{\sigma_0}^2}.
\label{eqn:rh}
\end{equation}
Based on its (questionably high) velocity dispersion (equation~\ref{eqn:sigma}), its $\mathcal{\widehat{M}}_\bullet$ black hole mass estimate (equation~\ref{eqn:Mhat}), and distance, we obtain $r_h = 17.7_{-12.2}^{+40.8}\,{\rm mpc} = 255_{-176}^{+591}\,\mu{\rm as}$ for NGC~3319*.\footnote{Because we question the discrepantly high $\sigma_0$ value from equation~\ref{eqn:sigma}, we alternatively can use the mass prediction of $\mathcal{\widehat{M}}_\bullet = 4.14_{-0.49}^{+0.50}$ (which does not consider equation~\ref{eqn:sigma}) to predict $\sigma_0$ from the $M_\bullet$--$\sigma_0$ relation. Reversing the relation from \citet[][equation~2]{Sahu:2019b}, we find that $\sigma_0=46.8\pm16.9\,{\rm km\,s^{-1}}$. Using this value now instead of the observed $\sigma_0$, equation~\ref{eqn:rh} yields $r_h = 27.4_{-16.5}^{+79.6}\,{\rm mpc} = 395_{-237}^{+1151}\,\mu{\rm as}$ for NGC~3319*.}

The Atacama Large Millimeter Array (ALMA) is useful for probing the gaseous cores of galaxies, including the rotating, torus-shaped, circumnuclear rings of molecular gas that enable measurements of the central black hole mass \citep[e.g.][]{Garcia-Burillo:2014,Yoon:2017,Combes:2019,Davis:2020}. ALMA currently has an impressive FWHM spatial resolution of 20\,mas at 230\,GHz. The East Asian VLBI Network \citep[EAVN; see][]{Wajima:2016,Hada:2017,An:2018} has achieved a spatial resolution of 0.55\,mas (550\,$\mu$as) at 22\,GHz. Similar milliarcsecond-scale resolution can be expected from the Long Baseline Array \citep[LBA;][]{Edwards:2015} and the European VLBI Network \citep[EVN; e.g.][]{Radcliffe:2018}. The Very Long Baseline Array (VLBA) could likely resolve the SOI of NGC 3319*, with its spatial resolution of 0.12\,mas (120\,$\mu$as) by utilising its longest baseline at 3\,mm, currently between Mauna Kea, Hawaii and North Liberty, Iowa.\footnote{\url{https://science.nrao.edu/facilities/vlba/docs/manuals/oss/ang-res}} The Event Horizon Telescope (EHT) can also resolve the SOI of NGC 3319*, with its PSF FWHM of 20\,$\mu$as. The EHT was able to resolve the emission ring, showing the event horizon, surrounding the SMBH M87* with a diameter of $42\pm3\,\mu{\rm as}$ \citep{EHT}.

Due to the difficulty of obtaining a direct measurement of the mass of NGC~3319*, it would be prudent to first study the AGN in NGC~3319 via reverberation mapping (RM) methods. In this respect, the bulgeless spiral galaxy NGC~4395 is the prototype. NGC~4395 possesses one of the least massive nuclear black holes that has ever been measured via direct methods. \citet{Brok:2015} obtained a black hole mass estimate of $4.0_{-1.0}^{+2.7}\times10^5\,\mathrm{M_{\odot}}$ via gas dynamical modelling; \citet{Brum:2019} similarly obtained $2.5_{-0.8}^{+1.0}\times10^5\,\mathrm{M_{\odot}}$ via gas kinematics. These direct measurements were preceded by informative RM black hole mass estimates of $(3.6\pm1.1)\times10^5\,\mathrm{M_{\odot}}$ \citep{Peterson:2005} and $(4.9\pm2.6)\times10^4\,\mathrm{M_{\odot}}$ \citep[][see also \citealt{Cho:2020} and \citealt{Burke:2020}]{Edri:2012}. Likewise, NGC~3319* could greatly benefit from further study by RM campaigns, or at least single-epoch spectra mass estimates.

\subsection{Implications}

The abundance, or scarcity, of black holes in this new mass domain of IMBHs, has a broad array of implications.  These include:
\begin{itemize}
\item Using low-mass AGN to extend the black hole scaling relations for predicting black hole masses in galaxies with quiescent low-mass black holes.
\item IMBHs will enable further refinement of the $M_\bullet$--$M_{\rm bulge,\star}$ and $M_\bullet$--$M_{\rm gal,\star}$ diagrams \citep[e.g.][]{Davis:2018,Davis:2019,Sahu:2019}, further facilitating the advancement of BH/galaxy coevolution theories \citep[e.g.][]{Kauffmann:2000,Croton:2006,Schaye:2015}.
\item Establishing the black hole mass function from stellar to SMBHs, and then revising the black hole mass density of the Universe should IMBHs prove abundant \citep{Aller:2002,Graham:2007,Shankar:2009,Davis:2014,Mutlu-Pakdil:2016}.
\item Increased understanding of the build-up of galaxies in our hierarchical Universe via merger events, including IMBH mergers; searches for and constraints of merger rate densities for IMBH binaries \citep{Salemi:2019,Jani:2019}.
\item Connections with nuclear star clusters and ultra-compact dwarf galaxies \citep{Graham:2009,Neumayer:2012,Georgiev:2016,Nguyen:2018,Graham:2019d,Neumayer:2020} and predictions for space-based gravitational wave detections involving longer wavelength gravitational radiation than ground-based interferometers can detect \citep{Portegies_Zwart:2007,Mapelli:2012,Fragione:2020}. The much-anticipated Laser Interferometer Space Antenna \citep[LISA;][]{LISA:2017} will have a designed observational requirement of detecting the coalescence of unequal mass black hole binaries of total intrinsic mass $10^4$--$10^6\,\mathrm{M_{\odot}}$ at $z<3$. The merging of such black holes (similar to NGC~3319*), each embedded in their nuclear star cluster, should coalesce within a Hubble time due to dynamical friction \citep{Ogiya:2019}. LISA and the next generation of gravitational wave observatories should also be able to find IMBHs in Milky Way globular clusters and the Local Volume \citep{Arca-Sedda:2020}.
\item The violent tidal disruption of white dwarf stars by IMBHs can trigger calcium-rich supernovae, spurring the nucleosynthesis of iron-group elements, and are capable of generating observable electromagnetic and gravitational-wave energies \citep{Rees:1988,Haas:2012,MacLeod:2016,Andreoni:2017,Andreoni:2019,Kuns:2019,Anninos:2019,Malyali:2019}.
\end{itemize}
IMBHs represent the grail lemma, needed to fill the void in our demographic knowledge of black holes, and tie up our inadequate theoretical understanding of BH/galaxy coevolution, feedback, and the growth of the Universe's most massive black holes. Increased future study of NGC~3319* promises to yield direct confirmation of the existence of an IMBH in AGN mode and offer immediate and lasting scientific advancement.

\section*{Acknowledgements}

We are grateful to Jonah S.\ Gannon, who provided valuable expertise with the spectroscopic analysis. BLD thanks David Nelson for the use of his secluded office space during the COVID-19 pandemic. This research was supported by the Australian Research Council's funding scheme DP17012923. Parts of this research were conducted by the Australian Research Council Centre of Excellence for Gravitational Wave Discovery (OzGrav), through project number CE170100004. This material is based upon work supported by Tamkeen under the NYU Abu Dhabi Research Institute grant CAP$^3$. This research has made use of NASA's Astrophysics Data System, and the NASA/IPAC Extragalactic Database (NED) and Infrared Science Archive (IRSA). We acknowledge the use of the HyperLeda database (\url{http://leda.univ-lyon1.fr}). We made use of the DS9 visualization
tool \citep{Joye:2003}, part of NASA's High Energy Astrophysics Science Archive
Research Center (HEASARC) software.

\bibliographystyle{pasa-mnras}
\bibliography{bibliography.bib}

\end{document}